\documentclass[useAMS,usenatbib]{mn2e}

\usepackage{graphicx}
\usepackage{txfonts}

\usepackage{natbib}

\newcommand{\mnras}[1]{MNRAS}
\newcommand{\apj}[1]{ApJ}
\newcommand{\apjl}[1]{ApJL}
\newcommand{\apjs}[1]{ApJS}
\newcommand{\aj}[1]{AJ}
\newcommand{\aap}[1]{A\& A}

\title[Structures in the fundamental plane]{Structures in the fundamental plane of early-type galaxies}
\author[D. Fraix-Burnet et al]{D. Fraix-Burnet$^{1}$\thanks{E-mail:
fraix@obs.ujf-grenoble.fr (DFB)}, M. Dugu\'e$^{1}$, T. Chattopadhyay$^{2}$, A.K. Chattopadhyay$^{3}$ and E. Davoust$^{4}$\\
$^{1}$Universit\'e Joseph Fourier - Grenoble 1 / CNRS, Laboratoire d'Astrophysique de Grenoble (LAOG) UMR5571, BP 53, F-38041 Grenoble cedex 9, France\\
$^{2}$Department of Applied Mathematics,  Calcutta University, 92 A.P.C. Road, Kolkata  700009, India\\
$^{3}$Department of Statistics, Calcutta University, 35 Ballygunge Circular Road, Kolkata 700019, India\\
$^{4}$Universit\'e de Toulouse, CNRS, Laboratoire d'Astrophysique de Toulouse-Tarbes, 14 av. E. Belin, F-31400 Toulouse, France}

\begin{document}

\date{Accepted May 26, 2010. Received ; in original form Decembre 2009}

\pagerange{\pageref{firstpage}--\pageref{lastpage}} \pubyear{2010}

\maketitle

\label{firstpage}

\begin{abstract}
The fundamental plane of early-type galaxies is a rather tight three-parameter correlation discovered more than twenty years ago. It has resisted a both global and precise physical
interpretation despite a consequent number of works, observational, theoretical or using numerical simulations. It appears that its precise properties depend on the population of galaxies in study.
Instead of selecting a priori these populations, we propose to objectively construct homologous populations from multivariate analyses. We have undertaken multivariate cluster and
cladistic analyses of a sample of 56 low-redshift galaxy clusters containing 699 early-type galaxies, using four parameters: effective radius, velocity dispersion, surface brightness averaged over effective radius, and Mg2 index. All our analyses are consistent with seven groups that define separate regions on the global fundamental plane, not across its thickness. In fact, each group shows its own fundamental plane, which is more loosely defined for less diversified
groups. We conclude that the global fundamental plane is not a bent surface, but made of a collection of several groups characterizing several fundamental planes with different thicknesses and orientations in the parameter space. Our diversification scenario probably indicates that the level of diversity is linked to the number and the nature of  transforming events and that the fundamental plane is the result of several transforming events. We also show that our classification, not the fundamental planes, is universal within our redshift range (0.007 -- 0.053). We find that the three groups with the thinnest fundamental planes presumably formed through dissipative (wet) mergers. In one of them, this(ese) merger(s) must have been quite ancient because of the relatively low metallicity of its galaxies, Two of these groups have subsequently undergone dry mergers to increase their masses. In the k-space, the third one clearly occupies the region where bulges (of lenticular or spiral galaxies) lie and might also have formed through minor mergers and accretions. The two least diversified groups probably did not form by major mergers and must have been strongly affected by interactions, some of the gas in the objects of one of these groups having possibly been swept out. The interpretation, based on specific assembly histories of galaxies of our seven groups, shows that they are truly homologous. 
They were obtained directly from several observables, thus independently of any a priori classification. The diversification scenario relating these groups does not depend on models or numerical simulations, but is objectively provided by the cladistic analysis. Consequently, our classification is more easily compared to models and numerical simulations, and our work can be readily repeated with additional observables.

\end{abstract}


\section{Introduction}

Physical understanding of astrophysical objects most often uses correlation diagrams. For early-type galaxies, such scaling laws have been for instance established on one hand between optical
luminosity and central velocity dispersion $\sigma$ \citep{FaberJackson1976}, and on the other hand between their effective radius $R_e$ and surface brightness averaged over effective radius $<\mu_e>$ \citep{Kormendy1977}.
These
correlations are rather tight, but the scatter is still reduced using a three-parameter relation of the form : $\log{R_e} = a \log\sigma + b <\mu_e> +\ c$
\citep{Dressler1987,Djorgovski1987}. This relation extends to faint and low-mass galaxies \citep[e.g.][]{Nieto1990}. This is the fundamental plane (hereafter FP) of early-type galaxies.

A long-running difficulty is the so-called ``tilt'' of the FP with respect to the ``virial plane'' obtained with the virial theorem and some simple assumptions about the population of early-type
galaxies.
Indeed, this tilt is different for different types of galaxies, like disk ones \citep[e.g.][]{Robertson2006,Hopkins2008}. Many studies have been devoted to this problem without a clear solution.
The
motivation is to obtain pure correlations both to constrain the models better and to use them as probes of characteristics which are difficult to measure or strongly biased. In particular, the FP
could in principle be a powerful tool to measure distances. But a proper calibration is required, and this appears difficult with its different tilts depending on the galaxy populations. 

Interpretation of the FP very often assumes some homology which is defined by \citet{Gargiulo2009} as: ``systems with density, luminosity and kinematics structures equal
over the entire early-type sequence and with constant mass-to-light ratios''. \citet{vanDokkum2003} define homology in such a way that the evolution of the FP is due only to the
evolution of $M/L$. These definitions are certainly linked to the assumption that all early-type galaxies are assembled in the same way, like dissipational mergers \citep{Robertson2006,Hopkins2008}. 

Reality is however more complicated \citep[see e.g.][]{Bender1992, Jorgensen1996, Borriello2003, vanderMarel2007} and early-type galaxies are very probably not all the result of the same
formation process. Basically, what is needed is some
invariant that allows us to trace a given object or class of objects through changes due to evolution. This invariant has been hoped to be the FP relation with some universality among a given
population, universality which is implied by the definition of homology above and characterized essentially by $M/L$. This provides a rough way of simplifying the many variables that
can evolve and may hide important sides of galaxy evolution. 

Models assuming only one homologous population have apparently failed to fit the FP in its entire extent. Selection criteria were proposed in order to obtain more
homogeneous samples and thus define what could be called a ``purer'' fundamental plane. However, these criteria are necessarily arbitrary, subjective and/or model dependent. The difficulty is
that many parameters are known to influence the global shape of the FP. Tilt of the FP \citep{Robertson2006}, warps, dispersion, changes with redshift, dependence of the mass-distribution on mass
\citep{Nigoche-Netro2009} among others, show that even though the FP looks tight
in the $\log\sigma$, $<\mu_e>$ and $\log{R_e}$ space, additional parameters could still play a role. 

The universality of the FP is also questioned: is it a bent plane \citep{Gargiulo2009} or a bent surface approximated by a collection of planes \citep{DOnofrio2008}? Obviously, this problem is related
to the choice of the sample, that is to the definition of homology. Theorists cope with many parameters that may influence the evolution of a given galaxy, especially when mergers are considered
\citep{Robertson2006}. Testing parameters one after the other, both theoretically and observationally, takes a lot of time, and might partly explain why after so many
years, this tight correlation keeps most of its mysteries. 

We think that it is time to explore new methodologies to better characterize and understand the FP relation. Multivariate clustering approaches are more objective in
selecting really multivariate ``homologous'' sub-populations of galaxies. This requires to explicitly assume that the ``fundamental plane'' is a priori not universal, and to understand it as
a
correlation in the $(\log{R_e}, \log\sigma,  <\mu_e>)$ space that could depend on the sub-population. Since galaxies are evolutive objects, homology can be more rigorously defined
by 'similarity due to same class of progenitor'. It is the astrophysical equivalent of 'similarity by common ancestry' in cladistics, a statistical method designed to relate evolutionary objects and
developed mainly by biologists. The use of many parameters is necessary to find the true homology, and to prevent analogy (same characteristics due to convergent evolution) to 
yield false lineages of galaxies. Cladistics does not \emph{assume a priori} properties linked to homology, but rather relies on
all the pertinent parameters to \textit{construct} homologous groups.

In the present paper, we have performed multivariate classifications with two independent approaches, cluster analysis and cladistic analysis, of a carefully chosen sample from the literature.
For this first kind of study, we consider only
the three parameters of the FP, $\log\sigma$, $<\mu_e>$, $\log{R_e}$, plus $Mg_2$ that are all given for this homogeneous sample. The first justification is that the number of parameters is
small, so it is not necessary to use PCA because we can physically discuss the variation in the light of these few parameters easily. The second advantage is that we are interested to study the
evolution of early-type galaxies in the light of their FP properties, and most of the authors have discussed evolution of galaxies on the basis of global FP.

The first approach we have used is by a multivariate technique known as K-means Cluster Analysis using the parameters above. The second approach, known as astrocladistics, is based on
the evolutionary nature of both galaxies and their properties \citep{FCD06,jc1, jc2,DFB09,FDC09}. The clustering technique compares objects for their
global similarities, while astrocladistics gathers objects according to their ``histories''. The two techniques are indeed complementary, the first one identifying coherent
groups, the second one establishing an evolutionary scenario among groups of objects. They are also totally independent, so that the comparison of their results is extremely instructive from a
statistical point of view.

This paper is organized as follows. The data and the different multivariate analyses we performed are described in Sect.~\ref{multivanal}. Comparison of the structures resulting from these
analyses, characterization of the fundamental planes of individual groups, evolution properties within the global FP and properties of the groups, are presented in Sect.~\ref{results}. The 
discussion on the meaning of the groups we have found is given in Sect.~\ref{discussion}, before our conclusion in Sect.~\ref{conclusion}. We present detailed descriptions of the cluster and
cladistic analyses respectively in Appendix~\ref{appendClus} and Appendix~\ref{appendClad}, and  provide additional diagrams in Appendix~\ref{otherfigures}.

\section{Multivariate analyses}
\label{multivanal}

\subsection{Data}

For this first study, exploratory, it was important to choose the sample carefully to be able to understand the result. We first need a homogeneous sample, devoid as much as possible of systematic biases. We need to remain at low redshifts, but ideally integrating several clusters. We are limited in the number of objects in the astrocladistic analysis. At the current stage of development of this novel approach, the reasonable limit is a thousand objects, and it is clear that a sample with $10^5$ objects is untractable directly.

We thus chose the data compiled and standardized by \citet{hudson2001} for 699 early-type galaxies in 56 clusters, in the redshift range from 0.007 to 0.053 (SMAC catalog). This sample has been carefully established in
order to obtain distance estimates from the FP and study streaming motions in galaxy clusters \citep{hudson2004}. The data include the three parameters of the FP space: $R_e$ in
arcsec,
$<\mu_e>$ in the R-band in mag arcsec$^{-2}$, $\sigma$ in km s$^{-1}$, plus the fully corrected magnesium index $Mg_2$ for nearly all of them. Distances were taken from the NED database ($H_0=73$ km/s/Mpc) to compute $\log{R_e}$ in
kpc. We performed analyses both with and without $Mg_2$ (hereafter 4- and 3-parameter configurations respectively). 
We use this parameter as a tracer of the abundance of light elements. While it increases as nucleosynthesis proceeds, it may not evolve at the same
pace as the metallicity (measured by Fe/H), because the light elements and Fe are produced in different kinds of supernovae arising in stars of different lifetimes.
However we are mainly interested in an indicator of chemical evolution, not in the detailed evolution of elements.
No error bars were used in this analysis. 

\subsection{Cluster analysis}

Cluster analyses do not accept missing values, so that they were performed with 696 galaxies in the 3-parameter
case (3 galaxies did not have redshift hence distance available), and 528 in the 4-parameter case (171 galaxies without $Mg_2$). 
The data for each parameter have been standardized to give them uniform weightage
as we want to classify the galaxies considering all the parameters as equally important.

With the K- means algorithms \citep{Hartigan1975}, we have obtained the optimum number of clusters given by the value of  K associated with
the largest jump and partitioned the galaxies according to this grouping (see Appendix~\ref{appendClus} for more details).

\subsection{Cladistic analysis}

We used the 699 galaxies for the cladistic analysis because for
missing values the algorithm simply guesses values that yield the most parsimonious diversification scenario. In this way, cladistic analysis also provides predictions for undocumented variables. 
The standardization is done automatically through the discretization of the variables (Appendix~\ref{appendClad}).

Even though the four parameters show mutual \emph{observational} correlations, they characterize a priori \emph{distinct} and  \emph{independent} physical properties that are expected to evolve and
for which some sort of evolutionary states could be defined. Note that this holds even when the variables are physically linked by the virial theorem, because the virial equilibrium is
a physical state that can evolve. These parameters are thus adequate for a cladistic analysis, in which case they are called ``characters''. 

The cladistic analysis yields trees from which groups can be defined. Groups on a cladogram are theoretically defined as evolutionary groups (or ``clades'') that comprise a node and all its
descending branches (see Sect.~\ref{structures}). The
identification of groups depends on the rooting of the tree, and must be analysed in regard with properties and other characteristics. The trees of the present study are rooted with objects or group
of objects having low $Mg_2$. 

Objects cannot reasonably be described as ``more evolved'' or ``less evolved'' than others because we do not have
an evolutionary clock. In addition, the measure of evolution is a very difficult concept in a multivariate space. This is why ``diversification'' is a more appropriate term. A very basic
illustration of this point on the hierarchical growth of dark matter halos can be found in \citet{DFB09}. See also Appendix~\ref{appendClad} and Sect.~\ref{evolFP} for some discussion on the measure
of diversification.

   \begin{figure}
   \centering
   \includegraphics[width=\columnwidth]{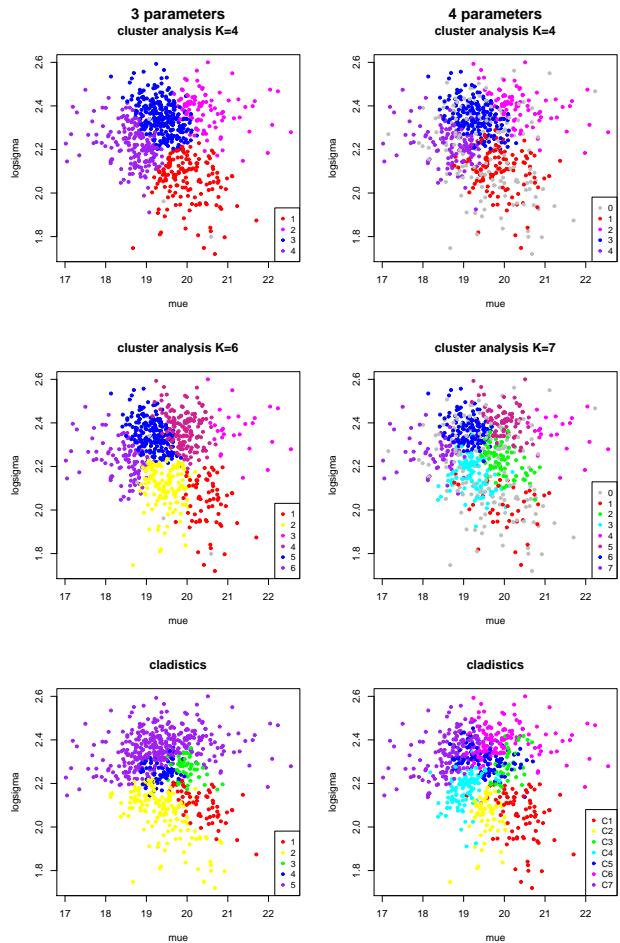}
 \caption{Groups obtained from cluster and cladistic analyses. Left column: with three parameters: $\log{R_e}$, $<\mu_e>$, $\log\sigma$. Right column: with the same three
parameters plus $Mg_2$. Top and middle row: cluster analysis results for the two possible numbers of groups as given by the K-means analysis in each case. Bottom row: cladistic results. The colour coding of groups is specific to each box. The colours for the cladistic analysis with 4 parameters (bottom right box) are the same as in Fig.~\ref{cladograms}. The grey colour of some points on the two upper panel to the right corresponds to missing $Mg_2$ data. Group number in the other plots is
used in Fig.~\ref{fighistocompgroups}. Group number ``0'' corresponds to galaxies that have missing $Mg_2$ values and were excluded from the cluster analyses with 4 parameters.}
    \label{gpclust}%
    \end{figure}
   \begin{figure}
   \centering
 \includegraphics[width=6 cm]{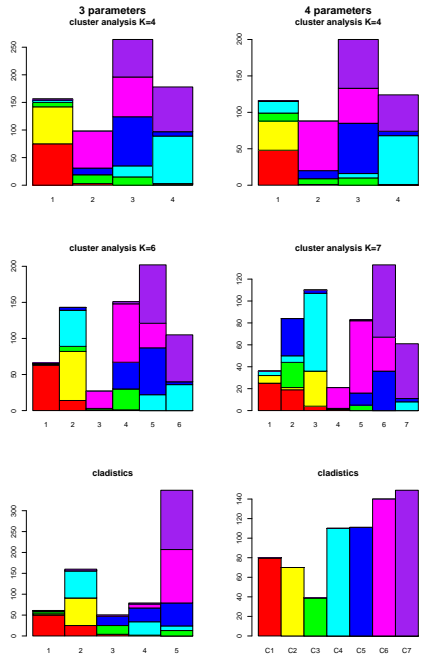}
   \caption{Histograms showing the distribution of the groups from cladistics with 4 parameters for each of the analyses. Same arrangement of plots as in Fig.~\ref{gpclust}. The colours of the groups are the same as in Fig.~\ref{cladograms}.}
    \label{fighistocompgroups}%
    \end{figure}

\section{Results}
\label{results}

\subsection{Structures in the fundamental plane}
\label{structures}

\begin{table*}
 \centering
 \begin{minipage}{16 cm}
  \caption{Main correspondence, drawn from Fig.\ref{fighistocompgroups}, between groups from cladistics with 4 parameters and the other groupings. In this table, contributions smaller than about 15\% are ignored. Note that some groups are roughly subdivided, like C6a, C6b.}
     \label{tabgroupcorresp}
  \begin{tabular}{llllllll} 
  \hline
                                                         & 1         &   2                   &  3                   &   4                  &  5                  &  6                     & 7 \\ 
	  \hline
4 parameters cladistics                    & C1       & C2                   & C3                  & C4                   & C5                & C6                    & C7 \\
4 parameters cluster analysis K=7   & C1a     & C1b+C3+C5a  & C2+C4           &  C6a                 & C6b             & C5b+C6c+C7a  & C7b \\
3 parameters cluster analysis K=6   & C1       & C2+C4a           & C6a               & C3+C5a+C6b   & C5b+C7a    & C4b+C7b           & --  \\
3 parameters cladistics                    & C1       & C2+C4a           & C3+5a            & C4b+C5b       &  C6+C7   & --      & -- \\
4 parameters cluster analysis K=4   & C1+C2& C6a                  & C5+C6b+C7a  & C4+C7b         &  --                  & --      & -- \\
3 parameters cluster analysis K=4   & C1+C2& C6a                  & C5+C6b+C7a  & C4+C7b         &  --                   & --      & -- \\
 \hline
\end{tabular}
\end{minipage}
\end{table*}

The results of our analyses are synthesized in Fig.~\ref{gpclust}, where the distribution of the groups in the $\log\sigma$ vs $<\mu_e>$ plane is presented, for the 3-parameter case on the left side,
and the 4-parameter case on the right side. The K-means analysis finds two peaks, corresponding to the two upper rows of Fig.~\ref{gpclust}, at K=4 and K=6 with 3 parameters (without $Mg_2$), and at
K=4 and K=7
with 4 parameters (with $Mg_2$). As described in Appendix~\ref{appendClus}, in both cases, the K=4 peak is lower than for K=6 or K=7. From a purely statistical point of view, the two latter numbers of
groups (middle row of the figure) are thus favoured. The cladistic result is shown on the bottom row of Fig.~\ref{gpclust}. Five groups can be distinguished on the tree obtained in the 3-parameter
case, 4 of them being very probably true evolutionary groups, the other one being an ensemble of successive branches. In the 4-parameter case, 7 groups are identified (Sect.~\ref{evolFP}).

It is obvious from Fig.~\ref{gpclust} that all our analyses agree to a small number of groups (4--7) and yield very similar groupings. In
particular, all the groups define separate regions on the plane, not across its thickness. It is important to keep in
mind that all the groupings mentioned in this paper have a statistical meaning, that is their borders are not
deterministic, and they are defined in a multivariate space so that overlapping can be important on 2D or 3D projections.

There is a very good similarity between the 3- and 4-parameter configurations for a given statistical method. For the cluster analysis, with K=4, there is nearly no difference between 3- and 4-parameter
results. For K=6 and K=7, the inclusion
of $Mg_2$ splits two of the groups (red and violet-red) into three (red, green and violet-red) along $\log\sigma$ and also slightly along $<\mu_e>$. The yellow group for K=6 is reduced in K=7
(cyan) because its low-$\log\sigma$ part have many objects with missing $Mg_2$ indices (represented as grey on Fig.~\ref{gpclust}) thus excluded from the 4-parameter analysis. For the cladistic
results, the difference appears to
lie along $<\mu_e>$ in the high-$\log\sigma$ part of  the plot (one group --purple-- split into two --magenta and purple-- groups) and along $\log\sigma$ for the yellow group that  is divided into
two (yellow and cyan). 

The two top rows of Fig.~\ref{gpclust} show that the cluster analysis tends to divide the fundamental plane in more or less equal
regions, suggesting a grid which is tilted with respect to the two axes $\log\sigma$ and $<\mu_e>$. Schematically, the four cases of the cluster analysis show the same structure: the high $\log\sigma$ part of the plot is divided in 3 or 4 groups, mainly but not exactly along $<\mu_e>$, while the low $\log\sigma$ part is divided between 1 and 3 groups. The dividing border between these two parts of the plots is more or less diagonal. This behaviour is particularly obvious with $K=4$, and the borders appear fuzzier in the case with 4-parameter and K=7.

This trend is still present but less obvious for the 4-parameter results (bottom row of Fig.~\ref{gpclust}). Noteworthy, adding $Mg_2$ does not reinforce any grouping trend along $\log\sigma$, that is the apparent grid is still tilted in the same direction, which could have
occurred in the cladistic analysis if the two characters were redundant. Borders are somewhat fuzzy, but the most notable difference is that the high $\log\sigma$ part of  the plot is divided into only 1 or 2 groups. The group C5 overlaps several other groups in the middle. These slight differences between the
two kinds of methods reflect the difference in the classification philosophy. But still, the distribution of the groups is essentially the same in all cases. In the high $\log\sigma$ part of  the plot,
the 4-parameter cladistic result is thus closer to the cluster analysis results than the 3-parameter one. In the low $\log\sigma$ part, the group C4 (cyan colour) of the 4-parameter cladistic analysis is nearly the same as the cyan group of the 4-parameter
cluster analysis with K=7. The group C2 (yellow) is composed of galaxies with missing $Mg_2$ indices that are not classified in the 4-parameter cluster analyses. From this point of view, the 4-parameter cladistic and cluster results are also quite similar.

A slightly more precise comparison is made in Table~\ref{tabgroupcorresp} for the principal correspondences between groups, and a more quantitatively one is presented in Fig.~\ref{fighistocompgroups}.
It must be noted that the group C5 is often
split in two parts that match nearly perfectly the two small subgroups visible in Fig.~\ref{cladograms} (see Sect.~\ref{evolFP}). 

Such an agreement between results obtained with two different techniques and two sets of parameters gives a high level of confidence in the structures of the FP of
early-type galaxies that we find. Since from a statistical point of view, the K=4 cases are less supported in the cluster analyses, we will not discuss them any further. Since
the 4-parameter cladistic result is slightly more robust than the 3-parameter cladistic analysis (Appendix~\ref{appendClad}), and it is also more similar to the
cluster analysis results especially with 4 parameters and K=7, we concentrate on the 4-parameter cases in the rest of this paper. Finally, since the cladistic analysis
additionally yields the
evolutionary relationships between groups, we will mainly discuss these 7 groups C1-C7 (Fig.~\ref{cladograms}) in the following. Nevertheless, to be complete and allow the reader to check that our interpretation is not limited to one specific analysis, but relevant for all of them, we present diagrams for the other groupings in Appendix~\ref{otherfigures}. 

   \begin{figure}
   \centering
 \includegraphics[width=\columnwidth]{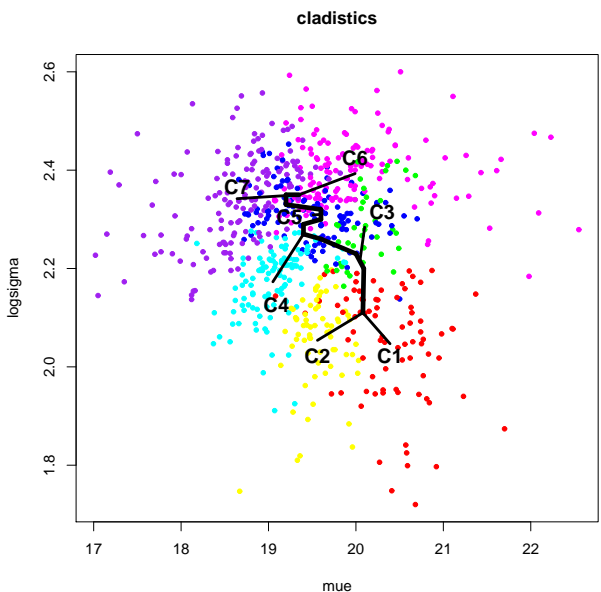}
   \caption{Projection of the tree in Fig.~\ref{cladograms} on the $\log\sigma$--$\mu_e$ plane. Each small branch leads to the group average given in Table~\ref{tablechar}.  The colours of the groups are the same as in Fig.~\ref{cladograms}.} 
    \label{evolrela}%
    \end{figure}

   \begin{figure*}
   \centering
 \includegraphics[width=16 cm]{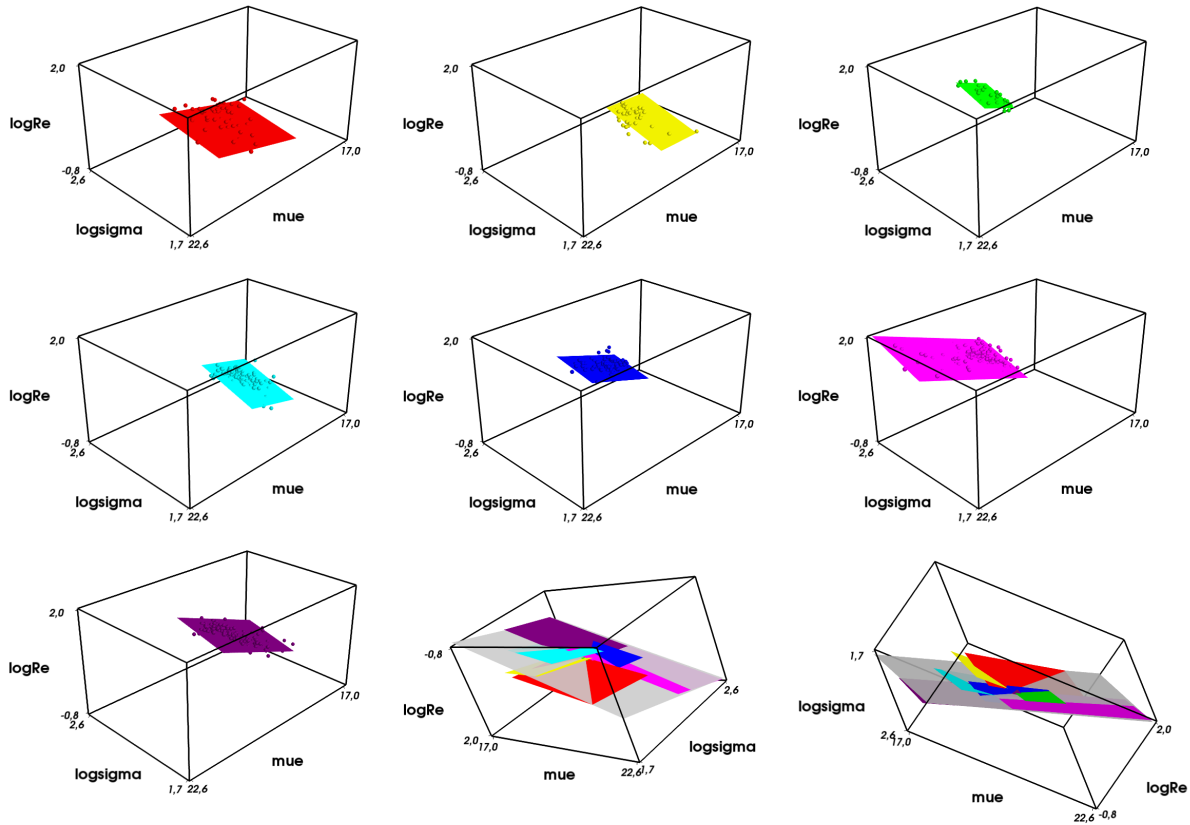}
   \caption{Three-dimensional views of the individual fundamental planes. The grey plane on the two bottom right plots is the global fundamental plane.  The colours of the groups are the same as in Fig.~\ref{cladograms}.} 
    \label{plans3D}%
    \end{figure*}

\subsection{Evolution within the fundamental plane}
\label{evolFP}

The groups from cladistics are defined from the most parsimonious tree found by the analysis. These groups are only a part of the information provided by the tree, that is shown in
Fig~\ref{cladograms} for the 4-parameter case. It depicts the evolutionary relationships between the groups, and has been rooted with C1 to orientate the direction of diversification,
that is the evolutionary distance from more primeval classes of objects (see Sect.~\ref{multivanal}). On the tree, at least 8 sets of branches can be seen, but only 7 are identified as groups for
this paper (C1 to C7). 

As shown by the
colours of the branches in Fig.~\ref{cladograms}, C1 has the lowest mean $Mg_2$. It has been chosen to root the tree (Sect.~\ref{multivanal}). Following the course of diversification, a split occurs
with a branch leading to C2 which is slightly more metallic.
The node at the split indicates a divergence in the evolution of some or all of the properties: C2 shares a common ancestorship with C3 to C7 -- low metallicity progenitors that could resemble C1
objects -- but has unique characteristics that we must identify in the multivariate space of evolutionary parameters. C3 to C7 are more diversified than C2 with respect to C1.

Very similar to C1, group C2 has a relatively low $Mg_2$ on average. These two groups seem to differ mainly in $<\mu_e>$. They both lie in the region where the global FP is
most distorted and dispersed (Sect~\ref{sectplanes}). Their FPs are very different from each other and from the other ones, both in orientation and thickness.

Next in the diversification, the small group C3  is very interesting because it appears rather early in the diversification scenario but is very similar to groups C6 and C7 in many respects
(see Sect.~\ref{sectplanes}, Sect.~\ref{propclades}, Table~\ref{planes}).

Group C4 is a well individualized group that is very robust since it appears nearly identical in all analyses (Fig.~\ref{fighistocompgroups}). It is grouped with C2 in two of the 3-parameter analyses, while on the tree, it is well separated from the latter, with C3 in-between. The discrimination between C2 and C4 thus appears when adding $Mg_2$ and is explained by the evolutionary relation or correlation of this parameter with the three other ones. Note that C2 has a large number of undocumented $Mg2$ values (29 out of 70, Table~\ref{tablechar}), which were not taken into account in the 4-parameter cluster analyses. This might explain why only C4 appears in all the analyses (Fig.~\ref{gpclust}).

Group C5 is not defined as an evolutionary group but by an ensemble of individual branches-galaxies and two small groups. Strictly speaking, group C5 is thus not  an
evolutionary group like the other ones. We decided not to individualize these two small
subgroups to avoid too much complexity in the diagrams, to stick as close as possible
to the results of the 4-parameter cluster analyses that find 7 groups, and because small subgroups could be hard to characterize due to a lower statistical significance. Even though we have
noticed that
these two subgroups can be clearly distinguished in some of the diagrams, their existence does not modify the main results of this paper. It is interesting to note that the two subgroups of C5 show up in Fig.~\ref{fighistocompgroups} and Table~\ref{tabgroupcorresp} as C5a and C5b. The first one is next to C3 in the low $\log\sigma$ part of the plot in Fig.~\ref{gpclust}, and the second one is within C6 and C7 in the high $\log\sigma$ part. This shows that we could have defined 3 groups in this region instead of 2, to be compared with the 4 groups of the 4-parameter cluster analysis with K=7.

As already mentioned, C6 and C7 are the two most diversified groups in our sample, but their respective places on the tree of
Fig.~\ref{cladograms} could be inverted without modifying any of our conclusions. They are simply two different groups, more similar to each
other than to other groups, and that are located at the end of our diversification scenario. There is also a subgroup in C7, but we do not think it deserves a
particular identification in the present paper.

\begin{figure}
   \centering
 \includegraphics[width=\columnwidth]{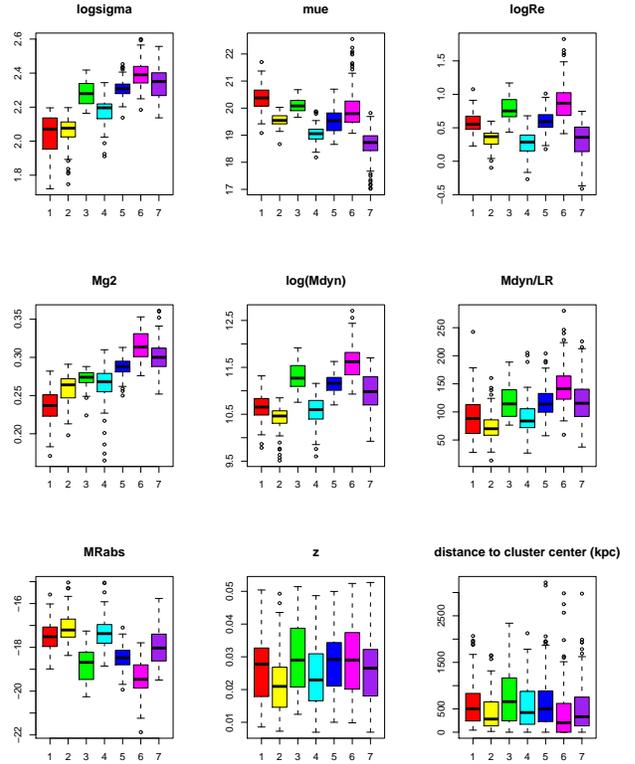}
   \caption{Statistics of 6 parameters for the groups from cladistics. Each boxplot for each group represents the 5 quartiles: median (thick bar at the middle of the box),
first and third quartiles (limits of the box), minimum and maximum (tick bars at the ends of the axis). Outliers are represented by circles. The colours of the groups are the same as in Fig.~\ref{cladograms}.} 
    \label{boxplot}%
    \end{figure}

The evolutionary scenario depicted by the tree in Fig~\ref{cladograms} can be projected on the global FP (Fig.~\ref{evolrela}). The thick track in Fig.~\ref{evolrela} represents the projection of the tree on the $\log\sigma$ --
$\mu_e$ plane, and the thin lines represent the beginning of the groups on the tree that are here manually extended up to the group average values (Table~\ref{tablechar}). Note that distances
apparent on this figure are only truncated measures of diversification which should be defined in the 4-parameter space. In this projection, the diversification within the global FP occurs with a main
trend along increasing $\log\sigma$ and some significant splits along $<\mu_e>$. 
The groups are roughly situated on each side of this trend, mainly along $<\mu_e>$ but also $\log{R_e}$ as shown on Fig.~\ref{boxplot}. C6 and C7 are clearly the two most
diversified groups, making a radical split, mainly in $<\mu_e>$, at the end of the main diversification trend.

The diversification within each groups in Fig.~\ref{evolrela} is rather complex, and we do not describe it in detail in this paper. Globally, diversification is always in a direction away from the main thick track.

\begin{table}
 \centering
 \begin{minipage}{\columnwidth}
  \caption{Fundamental planes for the 7 groups and the whole sample.obtained by linear regression. The equation of the fundamental plane is
expressed as $\log{R_e}=a \log\sigma + b\ <\mu_e> +\ c$. The last column gives the dispersion about the planes via the standard deviation of the perpendicular distance from the corresponding plane.}
     \label{planes}
  \begin{tabular}{lllll} 
  \hline
            &\hfil  a &\hfil b &\hfil c  & \hfil std \\
\hline
C1&$0.67\pm0.13$&$0.225\pm0.032$&$-5.4\pm0.8$&0.099\\
C2&$1.02\pm0.14$&$0.037\pm0.051$&$-2.5\pm1.0$&0.070\\
C3&$1.32\pm0.25$&$0.349\pm0.071$&$-9.3\pm1.2$&0.056\\
C4&$1.30\pm0.16$&$0.228\pm0.037$&$-6.9\pm0.7$&0.066\\
C5&$0.85\pm0.16$&$0.305\pm0.019$&$-7.3\pm0.6$&0.064\\
C6&$1.31\pm0.09$&$0.338\pm0.009$&$-9.0\pm0.3$&0.047\\
C7&$0.98\pm0.08$&$0.349\pm0.014$&$-8.5\pm0.2$&0.057\\
Full sample &$1.13\pm0.03$&$0.334\pm0.005$&$-8.5\pm0.1$&0.065\\
``virial plane'' & 2.0             & 0.4 &\hfil  -- & \hfil -- \\
\hline 
\end{tabular}
\end{minipage}
\end{table}

\subsection{Distinct fundamental planes for groups}
\label{sectplanes}

   \begin{figure}
   \centering
 \includegraphics[width=\columnwidth]{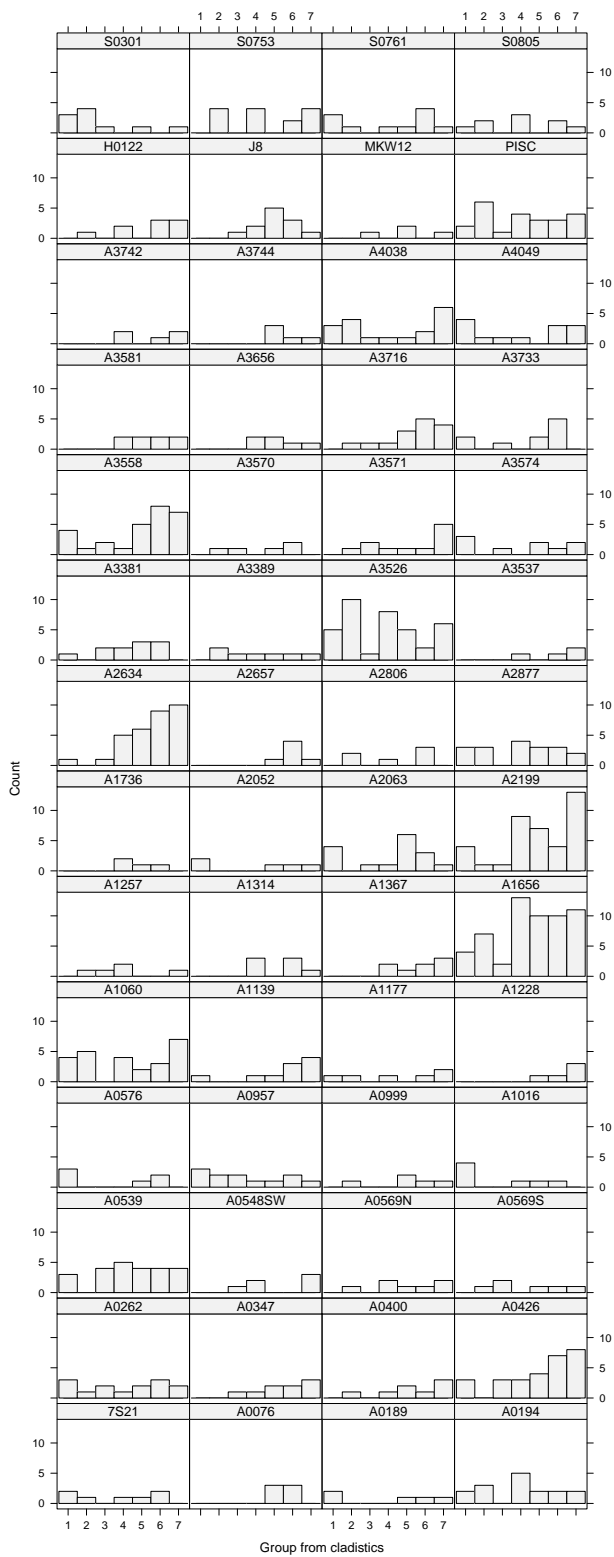}
   \caption{Histogram of group membership for each galaxy cluster. Note that the name of the cluster appears above each histogram.} 
    \label{histgalclust}%
    \end{figure}

In Table~\ref{planes} we report the coefficients found from a linear regression $\log{R_e}=a \log\sigma + b\ <\mu_e> +\ c$ done within each group and with the whole sample. The slope $a$ of the global FP is identical to that for group C2, compatible for groups C3, C4 and C6, significantly different for C7, and very different for C1 and C5. The slope $b$ for the global FP is identical to that for C6, barely compatible for C3, C5 and C7, and very different for C1 and C4. Group C2 is clearly peculiar (it seems to be independent of $<\mu_e>$). In summary, only C3 and C6 are very close to the global FP, the other ones differing more or less, either in one or the other projection or both. C3 and C6 are only very slightly closer to the virial plane ($\log{R_e}=2\ \log\sigma + 0.4\ <\mu_e>$) but not significantly. 

The dispersion about each plane -- the thickness of the plane -- is similar for the global FP, C2, C4 and C5, but much lower for C6, lower for C3 and C7, and very high for C1 (Table~\ref{planes}). The groups with tighter fundamental planes are not necessarily the most diversified since C3 is one of them. The two least diversified groups, C1 and C2, are well dispersed and certainly account to a large extent for the dispersion about the global FP.

Fig.~\ref{plans3D} presents a 3D view of the planes. Clearly the two less diversified groups C1 and C2, depart the most from the global FP. Groups C4 and C5 are also quite different. The two most diversified groups, C6 and C7, together with C3, have planes quite similar to the global FP. It is however difficult to see a progressive evolution in the respective orientations of the planes. We interpret this result as an indication that the global FP is not a bent surface, but made of a collection of several groups characterizing several fundamental planes with different thicknesses and orientations in the parameter space.

\subsection{Property distribution of the cladistic groups}
\label{propclades}

It is important now to understand physically why the groups have
been individualized by the multivariate algorithms. We again emphasize the fact that these groups have been identified in a multivariate space, and that it would be hopeless to try
characterizing them with one or two obvious properties. It is also important to remember that the classification comes from 4 independent, intrinsic and evolutionary pertinent parameters, so that
it can be enlightened by various properties even totally different from the one used in the multivariate analyses \citep[see our work on galactic globular clusters for a good example,][]{FDC09}. In particular, we have computed two quantities that are often used in studies of the FP: the dynamical mass $M_{dyn}$ and the $M_{dyn}/L_R$ ratio by using these two
relations: $<\mu_{e}>=-2.5\log(L_{R}/2\pi R_{e}^2) +4.45$  and $M_{dyn}\simeq A \sigma^{2}R_{e}/G$ with $A=3.8$ \citep{Hopkins2008}. It is important to note that the constant $A$ is
empirical and statistical for a given sample and is by no way universal. We then obtain: $\log(M_{dyn}/L_R) \simeq 2 \log\sigma -\log{R_e} + 0.4 <\mu_e> +\ C$, which is the ``virial'' plane if
 $\log(M_{dyn}/L_R)$ and $C$ are constant. The parameter $C$ depends on $A$ and some hidden physics, like dark matter, and is still not measurable on individual galaxies. We also computed $M_{Rabs}= - 2.5\log(L_{\odot R}) +4.45 = <\mu_e> - 2.5\log(2\pi r^2_e) +4.45$. Since we did not find B and R photometry for the full sample, we did not convert these quantities for the B band.

Average properties are given in Table~\ref{tablechar}, but boxplots are more convenient to visualize the distribution of some property within each group and evaluating differences between groups. Fig.~\ref{boxplot} shows that globally, groups have significantly different properties except for redshift and distance to cluster centre, the latter being very uncertain due to projection effects.

A trend toward increase with diversification is present for $\log\sigma$, $Mg_2$, and more marginally for $M_{dyn}$ and $M_{dyn}/L_R$. A trend toward decrease with diversification seems to be present for $<\mu_e>$ and marginal for $M_{Rabs}$. There is no trend for $\log{R_e}$. Note that the choice of C1 to root the tree of Fig.~\ref{cladograms} implies that the first group has the lowest $Mg_2$ on average, but it does not explain the regular trends because the relative positions of the groups are independent on this choice. 

Galaxies of group C1 have the lowest $Mg_2$, they are also large ($\log{R_e}$) relative to their mass $M_{dyn}$. They have a low surface brightness (high $<\mu_e>$). C1 and C2 show a low $\log\sigma$, but C2 galaxies are smaller, lighter and  brighter.

Galaxies of groups C3 and C6 are on average the largest, the most massive and the most luminous. These two groups however differ in $Mg_2$, and somewhat in $\log\sigma$ and possibly in the distance to the cluster centre. They have very thin FPs, close to the global FP (Sect.~\ref{sectplanes}). 

C4 galaxies have about the same mass and luminosity ($M_{Rabs}$) as the ones in C1 or C2, but they have a slightly higher $\log\sigma$ and surface brightness (lower $<\mu_e>$). C5 properties always lie exactly between C4 and C6.

Group C7, which also has a thin FP, is very different from C3 and C6. Its galaxies are as small as those of C4, with masses equivalent to the average mass of the whole sample. They have a high surface brightness (low $<\mu_e>$).

We find that all groups have very similar distribution of T-type morphologies (from -5 to -2, not shown), except C6 that has a strong excess of $T=-5$ galaxies.

As mentioned above, there is no relation between groups and redshift. There is no relation with galaxy cluster either: most clusters have objects from all groups (Fig.~\ref{histgalclust}). All have galaxies belonging to either C6 or C7 or both, with a tendency for the largest clusters to have more diversified objects. 14 clusters have no galaxies from C1 nor C2, but these are clusters with a small numbers of objects (less than 8 galaxies). Since we also found very similar results with subsamples composed of three well-populated galaxy clusters, we conclude that our classification is identical for all clusters.

We have performed a Kolmogorov-Smirnov test to check whether the distribution of group membership of the galaxies varies from cluster to cluster. Among the $C^2_{56}=1540$ possible pairs, only 66 (4\%) have a p-value lower than 0.05 (being equal to 0.012 for all the 66 pairs), indicating that there is evidence for a different distribution of membership in only 4\% of the cases. Interestingly, the 66 pairs all have one of the three clusters A0426, A0539 and A1656. The different distributions could indicate different environments and evolutionary histories for these clusters.

It appears that the diversification scenario cannot be described with only one or even two parameters. Our choice to root the tree with $Mg_2$ does not yield a perfectly regular increase of this parameter, and groups that are close in the diversification, like C6 and C7, show significant differences in most of the properties.

\begin{table*}
 \centering
 \begin{minipage}{140mm}
  \caption{Average properties of the seven groups of galaxies defined in Fig.~\ref{cladograms}. Ngal is the number of galaxies per group, NoMg2 is the number of missing $Mg2$ value in the data.}
     \label{tablechar}
  \begin{tabular}{llllllllll} 
  \hline
     &Ngal&NoMg2& $\log\sigma$    & $<\mu_e>$        & $\log{R_e}$   & $Mg_2$             & $\log(M_{dyn})$ &  $M_{dyn}/L_R$ & $M_{Rabs}$ \\ 
\hline
C1&80 &31&$2.047\pm0.114$&$20.39\pm0.45$&$0.58\pm0.16$&$0.235\pm0.024$&$10.63\pm0.31$&$ 90\pm38$&$-17.49\pm0.71$\\
C2&70 &29&$2.054\pm0.091$&$19.56\pm0.24$&$0.33\pm0.14$&$0.259\pm0.020$&$10.39\pm0.29$&$ 75\pm27$&$-17.06\pm0.69$\\
C3&39 &10&$2.284\pm0.074$&$20.10\pm0.26$&$0.78\pm0.19$&$0.271\pm0.014$&$11.31\pm0.31$&$120\pm32$&$-18.78\pm0.77$\\
C4&110&21&$2.173\pm0.072$&$19.05\pm0.30$&$0.28\pm0.17$&$0.265\pm0.026$&$10.57\pm0.29$&$ 89\pm29$&$-17.32\pm0.73$\\
C5&111&24&$2.309\pm0.054$&$19.51\pm0.46$&$0.60\pm0.16$&$0.288\pm0.013$&$11.16\pm0.19$&$119\pm29$&$-18.47\pm0.51$\\
C6&140&24&$2.393\pm0.079$&$19.99\pm0.72$&$0.87\pm0.27$&$0.316\pm0.017$&$11.60\pm0.35$&$147\pm35$&$-19.34\pm0.79$\\
C7&149&32&$2.342\pm0.091$&$18.64\pm0.54$&$0.33\pm0.25$&$0.301\pm0.020$&$10.96\pm0.40$&$119\pm36$&$-18.00\pm0.83$\\
\hline                     
\end{tabular}              
\end{minipage}             
\end{table*}               
                           
   \begin{figure}
   \centering
 \includegraphics[width=\columnwidth]{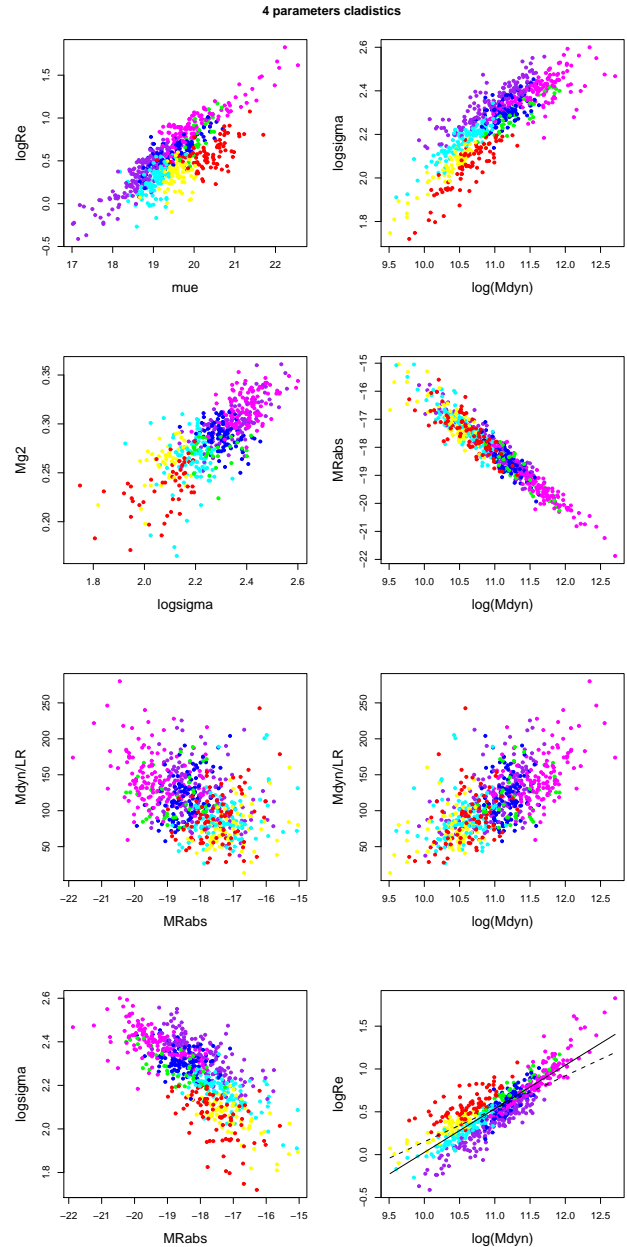}
   \caption{Scatterplots discussed in the text. The colours of the groups are the same as in Fig.~\ref{cladograms}. Note that
the values of $Mg_2$ for the C2 group (yellow) are predictions from the cladistic analysis (see Sect.~\ref{results}). In the bottom right diagram, the full line is a linear fit (slope of 0.51)  of the whole sample and the dashed line has a slope of 0.45/1.17 as found by 
\protect\citet{Robertson2006} for pure disk merger remnants (see text for details).} 
    \label{figcorr1}%
    \end{figure}

  \begin{figure}
  \centering
  \includegraphics[width=\columnwidth]{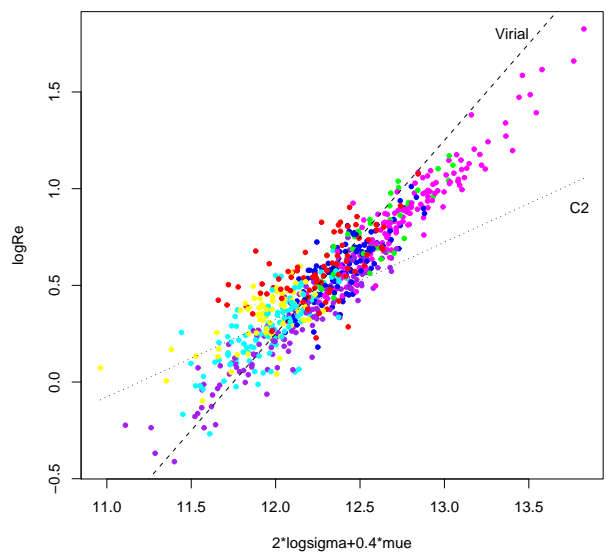}
\caption{Diagram showing $\log{R_e}$ as a function of $2*\log\sigma+ 0.4*\mu_e$ often used to represent the tilt of the FP with respect to the virial plane (dashed line). The dotted line is the fit for C2 which has the lowest slope of all groups. The slope tends to increase with diversification (see Table~\ref{tabRvsM}).}
   \label{figFPtilt}%
   \end{figure}

\begin{table*}
 \centering
 \begin{minipage}{160mm}
  \caption{Correlation coefficients for Fig.~\ref{figcorr1}. p-values are indicated in parentheses only when it is higher than $0.001$.}
     \label{tablecorrchar}
 \begin{tabular}{lllllllll} 
  \hline
                                                       &    all              &  C1                &  C 2                 &  C3              &   C4                 &  C5               &  C6                  & C7     \\ 
  \hline
$\log{R_e}$ vs $<\mu_e>$             &      0.59        & 0.24             &   0.03  (0.2)      & 0.55            & 0.34               & 0.63              &   0.79             &    0.80 \\
$\log\sigma$ vs $\log(M_{dyn})$     &      0.72        &  0.75            &    0.88              &  0.85           &  0.81              &  0.31             &    0.47            &     0.78 \\
$Mg_2$ vs $\log\sigma$                   &      0.59       &  0.23             &   0.24  (0.001)  & 0.01  (0.6)  & 0.008  (0.4)   & 0.08  (0.008) &   0.23             &    0.53  \\
$M_{Rabs}$ vs $\log(M_{dyn}) $    &      0.91        & 0.64             &   0.69                & 0.87            & 0.79              & 0.73               &   0.92             &    0.90  \\
$M_{dyn}/L_R$ vs $M_{Rabs}$       &      0.16        & 0.08 (0.01)   &  0.09  (0.01)     & 0.007  (0.6) & 0.08  (0.003)& 0.15               &  0.05  (0.009) &   0.11  \\
$M_{dyn}/L_R$ vs $\log(M_{dyn})$ &      0.42        & 0.11 (0.003) &  0.07  (0.03)     & 0.08  (0.09) & 0.03  (0.07)  & 0.02  (0.1)      &  0.24              &   0.37   \\
$\log\sigma$ vs $M_{Rabs}$           &      0.54       &  0.26             &   0.39                & 0.57            & 0.44              & 0.08  (0.002) &   0.36              &    0.56 \\
$\log{R_e}$ vs $\log(M_{dyn})$     &      0.73        &  0.53             &   0.79                & 0.90            & 0.87              & 0.69               &   0.85              &    0.89  \\
\hline                     
\end{tabular}              
\end{minipage}             
\end{table*}

\subsection{Property correlations of the cladistic groups}
\label{corrpropclades}

Two-dimensional scatterplots are useful to search for correlations, event though they are always projections of a multivariate space that naturally induce an apparent dispersion. We show several scatterplots in Fig.~\ref{figcorr1} and give the correlation coefficients in Table~\ref{tablecorrchar} for each group. It appears that correlations generally differ between groups and the whole sample.

Some correlations exist for both the whole sample and the groups individually. This is the case for the Kormendy relation between $\log{R_e}$ and $<\mu_e>$ that shows an important dispersion for the whole sample. However, it is particularly tight for groups C6 and C7 and rather weak for C1 and C4. There is no dependence of $\log{R_e}$ on $<\mu_e>$ for C2. It can be concluded that this correlation is not universal and only holds for well diversified galaxies. 

The relation between $M_{dyn}$ and $\log\sigma$ is also strongly dependent on the group (Fig.~\ref{figcorr1} and Table~\ref{tablecorrchar}). The correlation is generally very tight, significantly more than for the whole sample, except for C5 and C6 which show an important dispersion. It is important to recall that our computation of $M_{dyn}$ makes it essentially proportional to $\sigma^{2}R_{e}$, the coefficient possibly varying somewhat from galaxy to galaxy or from groups to groups \citep{Bolton2007,Hopkins2008}. 

Other correlations disappear within each group. The long-known global correlation between $Mg_2$ and $\log\sigma$ appears with a correlation coefficient of 0.59
(Table~\ref{tablecorrchar}) with the sample of the present paper. This
correlation is similar in C7 but much weaker in C1, C2 and C6. It even totally disappears for C3, C4 and C5. However, all groups are aligned according to our
evolutionary scenario of Fig.~\ref{cladograms}, the group C6 appearing before C7 (as already mentioned, C6 and C7
can be inverted in Fig.~\ref{cladograms} without modifying any conclusion). This result shows that the
correlation between $Mg_2$ and $\log\sigma$ is historical or spurious and not physical: it is the diversification itself that follows this track.

This is also true for the correlation between $\log(M_{dyn})$ and $M_{Rabs}$ which is very tight, with similar correlation coefficients for the whole sample, C3, C6 and C7. However,
the correlation is also largely defined by the succeeding groups along the correlation, C6 being the farthest at one end, and C2 and C4 at the other end, with C3 is between C7 and C6. 

Similarly, $M_{dyn}/L_R$ is roughly correlated to $M_{Rabs}$ and $M_{dyn}$, but more importantly the relation between these two parameters strongly depends on the
group. Indeed the first correlation is always weak, being slightly higher for the whole sample, C5 and C7. The groups seem to be arranged along the trend according to our diversification scenario, this trend delineating the global correlation. The second correlation between $M_{dyn}/L_R$ and $M_{dyn}$ is not very strong but still much tighter for the whole sample than for individual groups. The succession of the groups is present as well.

The Faber-Jackson correlation between $M_{Rabs}$ and $\log\sigma$ is a mixed case. It clearly depends on the group, being similar for the whole sample, C3 and C7, but
much weaker for the other groups. This correlation also roughly holds for the centroids of the groups which appear aligned along the global correlation according to their diversification rank.


A well known global correlation is found between $\log{R_e}$ and $M_{dyn}$, with a high significance (Table~\ref{tablecorrchar}) and a slope of 0.51 (Table~\ref{tabRvsM}, see Sect.~\ref{discussion}).
If we now look at individual groups, it is striking to see that they all
follow a tighter linear relation, except for C1 that is slightly more dispersed. The correlation coefficient for C5 is equivalent to that of the whole sample and very high for C3, C4, C6 and C7. Even more remarkably, they are very well separated from each other (Fig.~\ref{figcorr1}). Groups C1, C2, C4 and C7 are stacked perpendicularly to the correlation in an order that closely
follows the diversification scenario, going basically from the low masses and high radii toward higher masses and lower radii (see also Table~\ref{tabRvsM}). Groups C3 and C6 are displaced toward the
upper right of the correlation (high mass and high radius), and C5 is in an intermediate position. Each group occupies a finite region in the plot, both along and across the correlation. Since this
result is important, we show in Appendix~\ref{otherfigures} that it is true for all groupings found in the present work. Finally, the
slopes of the individual correlations (Table~\ref{tabRvsM}) are lower for C1 and C2, and then clearly increase with diversification being always larger than the global slope.

\section{Discussion}
\label{discussion}

\begin{table}
 \centering
 \begin{minipage}{\columnwidth}
  \caption{$\log{R_e}$ as a function of $2*\log\sigma+0.4<\mu_e>$ and $\log{M_{dyn}}$. The second column gives the slope of the linear regression fit to $\log{R_e}$ vs $2*\log\sigma+0.4<\mu_e>$ which measures the tilt of the fundamental plane with respect to the
virial plane. The
third column gives the slope $\alpha$ of the correlation: $\log{R_e}$ vs $\log{M_{dyn}}$ shown in Fig.~\ref{figcorr1}. Coefficients for dissipationless and dissipational systems comes from \protect\citet{Robertson2006} corrected for $M_{dyn}\propto M_{star}^{1.17}$. The two last columns give the median values for $\log{R_e}$ and $\log{M_{dyn}}$ in the corresponding group.}
     \label{tabRvsM}
  \begin{tabular}{ccccc} 
  \hline
   Group            & tilt / virial   & slope $\alpha$ & median       &   median        \\
                         & &        & $log(Re)$ & $log(M_{dyn})$  \\
\hline
C1                    &0.42& 0.38  &0.55&10.65\\
C2                    &0.40& 0.42  &0.37&10.47\\
C3                    &0.74& 0.56  &0.74&11.27\\
C4                    &0.61& 0.55  &0.29&10.61\\
C5                    &0.71& 0.69  &0.59&11.17\\    
C6                    &0.81& 0.71  &0.85&11.60\\
C7                    &0.70& 0.60  &0.36&10.98\\
Full sample       &0.70& 0.51  &0.51&10.99\\
dissipationless    &0.96&0.38  &&\\
dissipational       &0.83&$\simeq 0.49$ &&\\
\hline
\end{tabular}
\end{minipage}
\end{table}

\subsection{Structures in the fundamental plane space}

The first important result of this paper is that we find structures in the global FP of this sample despite that it was established to be very homogeneous. FP properties are known to differ from
one
sample to another, depending on kinds of objects and redshifts. They mostly differ in relative shift and somewhat in inclination. This implies that the global FP does not relate galaxies of the same kind, but rather traces a diversification scheme of
various groups of homologous galaxies. A priori, the classification we find is by no means universal. It is only valid for the sample used described with the 4 parameters $\log{R_e}$, $<\mu_e>$,
$\log\sigma$ and $Mg_2$. Other analyses have to be performed on other samples and results compared. We must mention that we find the same kind of result with a totally distinct sample of 500 nearby
early-type galaxies (Fraix-Burnet et al., in prep).

\citet{DOnofrio2008} have performed an analysis of the FP for 59 galaxy clusters with redshifts between 0.04 and 0.07, quite similarly to our study (4 clusters are in common: A0548SW, A2657, A3558,
A3716). They compare the FPs obtained for each cluster and find that they are not compatible with a universal FP. However, they find that the FP
properties depend strongly on the sample analysed. They use different subsamples, and find that probably the different ways early-type galaxies were selected have a prominent influence on the fit of
the FP. This could have serious consequences on the physics derived from the FP properties, like the structure of dark matter \citep{Borriello2003}. In our point of view, any a
priori selection criterion induces an unavoidable bias, even if enough care is taken for homogeneity or completeness. A multivariate classification is the best objective tool to correlate
clusters with FP properties and investigate whether the FP can be universal. In our approach, the groups we have found solely depend on the intrinsic properties of the galaxies, independently
of cluster or type selection. It appears that they do not correlate with cluster (Fig.~\ref{histgalclust}) and the distributions are identical in 96\% of the cases (Sect.~\ref{propclades}). This implies that both the populations of galaxies and the environments in our clusters do not differ much and the evolution of galaxy properties, that is the diversification of galaxies, is similar for nearly all clusters within our redshift range (0.007 -- 0.053). This also implies that our classification, not the FPs, is universal within these limits. 

\citet{Gargiulo2009} argue for a bent plane instead of a continuous ``distribution'' of planes depending on some arbitrary
pre-selection criterion (size, mass, type, ...). \citet{DOnofrio2008} find that the plane is actually a bent surface ``approximated by different planes depending on the
different regions of the FP space occupied by the galaxies of the samples under analysis.''. Our result is clearly against a universal bent plane or surface, because even a homogeneous sample like
the one we used, appears to be divided into several planes of slightly different inclinations. More significantly, our cluster analysis and cladistic results show that the multivariate space is
divided into different regions, related by the diversification process, in each of which a correlation might or might not exist. Somewhat in agreement with \citet{DOnofrio2008}, but for different reasons, we doubt that finding universal and unbiased coefficients for the FP is a realistic objective. We conclude that what is called the global FP is indeed a collection
of regions in a multivariate space, with projections on the corresponding 3-D space more or less planar, depending on the corresponding group. Each group can be considered as truly 
``homologous', as defined in the introduction (``similarity due to same class of progenitor'') because  it results from a multivariate analysis, not from arbitrary selection criteria.

Even when the FP is carefully built, a significant dispersion remains that suggests another parameter is at play. For instance, it has been found that this dispersion could be explained by the
mass-distribution depending on mass \citep{Nigoche-Netro2009}. Even though our determination of the mass is based on
the virial theorem, our result seems in agreement with this interpretation. Fig~\ref{boxplot}shows  that the variance within a group is larger for groups of more massive galaxies (C3 and C6). 
Unfortunately, mass, like most of physical parameters, is very difficult to measure \citep[see a discussion in][]{Hopkins2008}. However, we think that the dispersion about the FP is naturally due
to cosmic variance within homologous populations. In our approach, we conclude that mass-distribution depends on groups, and not necessarily and only on mass.
We note that for essentially all the parameters considered in this paper, we find that intragroup variance is often larger than intergroup differences (e.g. Fig~\ref{boxplot}). This is certainly intrinsic to the nature of
galaxies, and obviously calls for multivariate classifications. In particular, this shows that making binary classifications (like low- and high-masses objects) may introduce an important bias
physically unjustified.

The variation of the ratio $M/L$ with $z$ (i.e. due to passive evolution) is often used to explain offsets with respect to the FP that are observed at higher redshifts \citep[e.g.][]{vanDokkum2003}. This is different from the tilt which can be explained by a variation of $M/L$ with $L$. But assuming $M/L$ to be constant (with a given tilt) for a given redshift implies a very strong physical and evolutive assumption. 
This parameter is computed by using the three
parameters of the FP space plus the essential assumption of a constant ratio between dynamical mass and real mass for all galaxies. In this homogeneous sample used in our study, the ratio $M_{dyn}/L_R$
is not constant across and within the plane, and is different depending on the group (Fig.~\ref{boxplot} and Fig.~\ref{figcorr1}). In addition, we do not find any dependence of this parameter with $z$ but our range in redshift is
small. Our work
thus implies that $M/L$ is not constant for $z=0$ \citep[Sect.~\ref{propclades}; see also][]{Jorgensen1996}.

  \begin{figure}
  \centering
  \includegraphics[width=6cm]{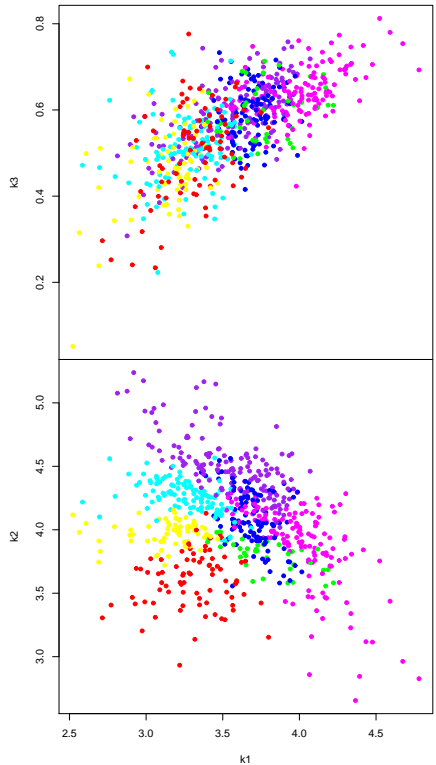}
\caption{Diagram showing our groups in the k-space as defined by \protect\citet{Bender1992}. The colours of the groups are the same as in Fig.~\ref{cladograms}.}
   \label{figk1k2k3}%
   \end{figure}

\subsection{Groups and assembly histories: the k-space}
\label{kspace}

\citet{Bender1992} introduced the k-space representation intended to depict the fundamental plane relation with more physical coordinates.
Our groups are represented in Fig.~\ref{figk1k2k3} and are outlined in the $k2$-vs-$k1$ projection. Note that our $k2$ and $k3$ axes depend on $\mu_e$ which is in the $R$-band, and thus appear shifted as compared to other diagrams made in the $B$-band.  \citet{deRijcke2005} have compared several samples of galaxies divided in fine morphological classes, partly using data from \citet{Bender1992}, and some semi-analytical models of galaxy formation and evolution more oriented toward dwarf galaxies. It is obvious on their $k2$-vs-$k1$ diagram that the distribution of groups in our classification appears in rather good agreement with their morphological classification. This is not surprising since 2 of our 4 parameters used in the cluster and cladistic analyses are structural ($\log\sigma$ and $\log{R_e}$), while a third one ($\mu_e$) is somehow correlated to stellar concentration. 

This agreement is interesting because our classification does not require very sophisticated image analysis procedures \citep[e.g.][]{Kormendy2009}. In addition, our classification is based on physical observed parameters that can be more easily compared to models and numerical simulations than detailed features in the images. Our multivariate analyses can also be extended to more observables to explore the complexity of galaxy diversification. 

Interestingly, the comparison of our $k2$-vs-$k1$ diagram with that from \citet{Bender1992} shows that our two groups C6 and C7 match rather well the distinct regions of respectively bright elliptical galaxies and bulges (of lenticular or spiral galaxies). Note that the data of \citet{Bender1992} are for the bulge component of galaxies almost exclusively SOs \citep[while][cite them as for ``bulges of spiral galaxies'']{deRijcke2005} whereas our values \citep[from][]{hudson2001} are for the entire galaxy. There are no spirals in the sample we use, but galaxies from C7 should somehow share a similar assembly history that these bulges. Groups C6 and C7 could be paralleled respectively to the two kinds of ellipticals suggested by \citet{Kormendy2009}: i) the giant ellipticals that are essentially non-rotating, anisotropic, triaxial, little flattened (E1.5), have cuspy cores boxy-distorted isophotes; ii) normal- and low-luminosity ellipticals that rotate rapidly, are relatively
isotropic, oblate-spheroidal, flattened (E3), coreless, have disky-distorted isophotes, most bulges of disk galaxies being like low-luminosity ellipticals. They also suggest that the first ones are formed from dissipationless (dry) mergers and the second ones by dissipative (wet) mergers. However, we do not believe that our dichotomy C6-C7 could be explained by a simple dichotomy in dissipation because such diversified objects have quite complex a history that cannot be summarized by one kind of assembly event (see below).

In particular, interactions are a quite frequent event that transform galaxies. \citet{Aguerri2009} have simulated the effect of rapid interactions on the evolutions of galaxies. We derive from their $k2$-vs-$k1$ plots that our groups, except for C1 and C2, seem to be robust with respect to these kinds of interaction, that is these events make a galaxy move within its group. This could mean that these groups cannot be formed by rapid interactions only. We note that their simulations do not have objects similar to C3 and C6 groups, but this might be due to the limited number of configurations they present. Anyhow, C3 and C6 galaxies are the most massive ones, implying very probably mergers and accretion to have taken place.

\subsection{Groups and assembly histories: tilt of the FP}
\label{assemblhistories}

The difference between the FP and the virial plane is often studied as a tilt in the $\log{R_e}$ vs $2*\log\sigma+0.4<\mu_e>$ representation. For instance, \citet{BoylanKolchin2005,Robertson2006} have performed
simulations of several scenarios of merging histories. They find that gas dissipation during mergers of disk galaxies can explain the tilt, even if \citet{Robertson2006} do not exclude other kinematical or photometrical effects.

If we compare the tilt of the FP for our different groups (Table~\ref{tabRvsM}) with their predictions, we find that our groups agree better with dissipational mergers with a lot of gas (coefficients of about 0.8) than with dissipationless ones (coefficient close to 1). Groups C3, C5, C6 and C7 are the closest to the virial plane, but still far from it. Groups C1, C2, C4, which are the furthest from the virial plane,
could either be mergers of systems with much more gas, or else have evolved by slow accretions or in isolation. Since
dispersion is significant within all groups and coefficients are much lower than those found by \citet{Robertson2006}, especially for less diversified ones, dissipation and gas fraction alone in
these merger scenarios cannot explain the differences between our groups. As mentioned in Sect.~\ref{kspace}, C1 and C2 may not be the result of mergers.

In addition, as we have shown in Fig.~\ref{plans3D}, the planes are tilted in the three directions of the FP space,
so that the correlation between $\log{R_e}$ and $2*\log\sigma+0.4<\mu_e>$ cannot be sufficient to understand all the physics of the complex histories of galaxies. \citet{Robertson2006} also give the variations of the plane coefficients $a$ and $b$ (see our Table~\ref{planes}) as a function of gas fraction. It appears that they are not monotonic and thus more difficult to interpret. Our coefficient $a$ is always much lower than that found by the simulations \citep[in the range 0.67 -- 1.3 as compared to 1.55 -- 2.07 in Table~3 of ][]{Robertson2006} and our coefficient $b$ in the lower range. 
However, they use the surface density of stellar mass, whereas we use the surface brightness density. The relation between the two could well distort the plane in the three directions, and this comparison is probably flawed. In addition, it seems that the slope $a$ varies much in the literature \citep{Bernardi2003} depending on the wavelength band and the fitting method. Finally, the most discrepant groups in $a$, C1, C2 are probably not the result of mergers so cannot be compared to such simulations.

\subsection{Groups and assembly histories: radius vs mass}

The relation between $R_e$ and $M_{star}$ is a good diagnostic tool for merger histories \citep[e.g.][]{BoylanKolchin2005,Robertson2006,Hopkins2009}. It seems that $M_{dyn}/M_{star}\propto M_{star}^\lambda$ with $\lambda\simeq 0.17$, but it depends on each galaxy so that this relation has a lot of scatter \citep[e.g.][]{Robertson2006}. Because of this unknown scatter, we here use $M_{dyn}\propto M_{star}^{1.17}$ in Fig.~\ref{figcorr1} and Table~\ref{tabRvsM}. 

We find a global correlation with a slope of 0.51, in agreement with the SDSS fit for early-type galaxies \citep[0.56/1.17=0.48][]{Shen2003} but steeper than that found in the simulation result for pure disk merger remnants \citep[0.45/1.17=0.38,][]{Robertson2006}. Dissipational systems (their figure 4 for merging of gas-rich disk galaxies with dark matter halos, star formation and supernova feedback) give steeper slopes, similar to those of groups C1, C2 but still not as steep as for all the other groups especially C5 and C6. 
More puzzling is that, according to \citet{Robertson2006} result, the bulge tends to diminish the slope, whereas we find steeper slopes for all the
groups except C1 and C2. However the steeper slopes might be due to mergers of non-disky objects or the result of repeated merging of small systems \citep{Shen2003}. 

Apart from the slope of the $\log{R_e}$ vs $\log{M_{dyn}}$ correlation, we see from \citet{Robertson2006} that the groups that are below the global relation have older progenitors, that is
progenitors with properties characteristic of higher redshifts. This would imply that groups C1 and C2 (assuming they are the result of a merger) have formed recently, hence they had not enough time to diversify much, while C7 is clearly the
remnants of higher redshifts progenitors. It is interesting to note that C6, which from several characteristics could be considered as more diversified than C7, seems to be the remnant of slightly more recent
progenitors. But since it occupies the high mass and high radius region of the plot, it is reasonable to think that its galaxies are the result of several mergers, the last ones being more recent
than for C7 objects. Similarly, C3 is above C6, showing that its galaxies had a more recent merger. This seems to contradict the high resemblance between these two groups with C3 having a lower $Mg2$ (Fig.~\ref{boxplot}), but metallicity is also governed by the composition of the merging galaxies.

Mergers with wide orbits are also situated above the global $\log{R_e}$ vs $\log{M_{dyn}}$ relation \citep{Robertson2006}, where C1, C2, C3 and
partly C6 lie. We could then conclude that groups C1 and C2 are closer to dissipationless remnants of pure disk wide-orbit mergers. Together with our discussion in Sect.~\ref{kspace} and Sect.~\ref{assemblhistories}, this seems to confirm the idea that these objects are the remains of interactions or monolithic collapse rather than solely mergers. C3 seems to lie above C6, suggesting mergers of wider orbits. This apparently contradicts its lower $\log\sigma$ (Fig.~\ref{boxplot}) but several different kinds of mergers probably occurred, with some particularly violent for C6 galaxies.

Other insights are obtained from \citet{Ciotti2007}. Indeed, the interpretation of their results should be reconsidered for each of our groups instead of the whole sample since the $R_e$ vs
$M$ plots show different slopes depending on the models. For instance, they find that 10-equal-mass parabolic mergers yield steeper slopes. An important result of their simulations is that parabolic dry mergers cannot explain the formation of the scaling laws of early-types, but wet mergers do, except maybe for very massive galaxies \citep[see also][]{BoylanKolchin2005,Hopkins2009}. Dry mergers however do not modify these relations once they exist. This means in particular that the FP is formed via wet mergers. Our groups C3, C6 and C7 show the thinnest FP and consequently should have formed in this manner. Since C3 is less metallic, this merger was more ancient. Then subsequent dry mergers have increased its mass to the level of C6, whereas C7 galaxies were not subject to important mass increase. We however cannot exclude that C3 might be the result of an important monolithic collapse with some subsequent low-metallicity dry mergers. These mergers were probably slightly less violent in C3 (having a lower $\log\sigma$), which could be explained by their average larger distance to cluster centres than C6 (Fig.~\ref{boxplot}).

In real observations, the scatter in the $\log{R_e}$ vs $\log{M_{dyn}}$  correlation is quite large \citep{Shen2003,Franx2008}. Indeed, it is is much larger for the whole sample than for each evolutionary group (Fig.~\ref{figcorr1} and Table~\ref{tablecorrchar}). The observational scatter is thus explained essentially by the ``stacked'' distribution of the groups. \citet{Franx2008} suggest that the velocity dispersion ($\sigma^{2}\propto M_{dyn}/R_{e}$) or the stellar surface density  ($\sim
M_{dyn}/R_{e}^2$) might explain the scatter of their correlation with colour. Unfortunately, we do not have access to colours for most of the galaxies of our sample. 
We find that $\log{R_e} =
\alpha*\log{M_{dyn}}+const.$ with $\alpha$ given in Table~\ref{tabRvsM} depending on the group. This naturally yields the relation $M_{dyn}\propto R_e^{1/\alpha}$ that is different from group to group. Consequently,
our result shows that the scatter in the $\log{R_e}$ vs $\log{M_{dyn}}$ diagram is mainly explained by the variation of the \emph{relation} between
mass and radius, this relation implying a physical interpretation only valid for a particular group. Since $1/\alpha$ ranges from 1.41 to 2.63 and is never close to 2 (except for the whole sample!), this seems to indicate that the neither velocity dispersion nor stellar surface density entirely explain the scatter in the relations between $\log{R_e}$, $\log{M_{dyn}}$ and colour. Thus our groups are probably not homogeneous in colours. 
We finally note that our Fig.~\ref{figcorr1} shows that averaging values of $R_e$ over bins of  $M_{dyn}$ as in \citet{Shen2003}, mixes different populations of
galaxies and consequently introduces an artificial scatter within each bin.

\subsection{Physical interpretation of the diversification scenario}

Even though the above comparisons are limited by the fact that numerical simulations could not consider all possible scenarios, they show that the groups found by cladistics are evolutionary groups, each one gathering objects with the same assembly history.
We here summarize a possible explanation for the origins of the different groups. It must be understood that the assembly and interaction histories of nearby galaxies are necessarily complex and comprises often several significant transforming events \citep[e.g.][]{jc2}.

 \begin{itemize}
\item C1 : is chosen as the most primeval group because of its low average $Mg_2$. Its galaxies are rather faint, and relatively large for their masses, with a low $\log\sigma$. They might be the remains of a simple assembly through a monolithic collapse with little dissipation, and they were probably perturbed by interactions. 
\item C2 ; its galaxies are less massive and smaller than the ones in C1, and they have a slightly higher $Mg_2$. They are also somewhat brighter. They could be the remains of wind stripping of some kind of more diversified objects.
\item C3 : large, massive and bright, they have a low metallicity $Mg_2$ with a high velocity dispersion but less than for C6. Although this is of low significance, they could be the most distant galaxies of cluster centres. They could be the remains of an ancient wet merger with subsequent low-metallicity dry mergers. We cannot exclude that C3 might be the result of an important monolithic collapse with some subsequent low-metallicity dry mergers. 
\item C4 : its galaxies look much like the ones in C1 and C2, but more concentrated, with a metallicity slightly higher than for C2. Are they simply galaxies in which stellar formation has been continuous, or are they C1 galaxies that were initially richer in gas that has not been swept like for C2 objects? They could also be the remnants of several minor mergers and accretion. 
\item C5 : they have intermediate properties between the galaxies of C4 and those of C6. This group is indeed composed of two small evolutionary groups, one closer to C4 on the FP and the other one more or less in-between C6 and C7. Are they intermediate objects? We think that they could be objects that cannot be fully described by only 4 parameters. Their apparently odd position within the FP would the be due to a projection effect of a larger multivariate space onto our 4-parameter classification.
\item C6 : its galaxies may be the closest to cluster centres, they define the tighter fundamental plane. All the parameters of Fig.~\ref{boxplot} are the highest of the sample except for $\mu_e$. Most of the galaxies are probably large ellipticals usually designated as Es. Their history might be complex, with many transformation events, they could represent a kind of end state of galaxy diversification. They might be the remains of many accretions, minor mergers, together with more or less dissipational major mergers. However, they seem to be the remains of both wet and dry mergers, the most recent ones being of the latter kind.
\item C7 : its galaxies also define a tight FP like C6, but they are the smallest of the sample, with the highest surface brightness $\mu_e$. They are very metallic. They seem to be associated with the remains of a dissipative (wet) merger, with very little or no dry mergers. They could also have formed through minor mergers and accretions. It is striking that they occupy the same region in the k1-k2 space as the bulges (of lenticular or spiral galaxies).
\end{itemize}

In view of the present results, we conclude that the FP relation depends on the group and thus on the histories of galaxies. Obviously, the tightness of the relation is different from one group to the other being much less for less diversified groups. For more diversified groups, the correlation is tight but still very significantly different from the virial plane. In a similar manner as for the bivariate plots of Fig.~\ref{figcorr1}, it is possible that the correlation could be historical and not physical. In other words, this planar correlation between $\log{R_e}$, $\log\sigma$ and $\mu_e$ might not be a tilted virial plane due to dissipation or a particular behaviour of say $M/L$, but rather a parametric correlation between the evolutions of these parameters. If this correlation is tighter in more diversified objects, this is probably because they are the result of a mixture of several or many transformation events like collapse, interaction and merging, dry and wet. To investigate this point further, our analyses should be redone with additional measured values of mass, luminosity and $M/L$ at least. Then a study of the diversification within each group should reveal the combined evolutions of all parameters.

\section{Conclusions}
\label{conclusion}

In this paper, we have reconsidered the study of the so-called fundamental plane of early-type galaxies. We have used two different multivariate clustering tools, cluster analysis and cladistic
analysis, to explore the 4-parameter space $\log\sigma$, $<\mu_e>$, $\log{R_e}$ and $Mg_2$. With both methods, we used both the three first observables, those of the fundamental plane
correlation, and the four altogether. 

The sample used in our analysis, taken from \citet{hudson2001}, has 699 objects spread into 56 galaxy clusters. In all the analyses, 4 to 7 groups are found,  which are consistent from both
statistical and physical arguments. The very good agreement between our
different analyses provides a good confidence in the existence of structures \textit{within} the FP. We emphasize that no a priori criterion is used to select groups of objects, even in the multivariate space.  In this paper, we focus on the 7 groups (labelled C1 to C7) defined by the cladistic analysis because it additionally provides a diversification scenario linking the groups. Note that since we are in a multivariate space, the wording ``diversification'' is more appropriate than ``evolution'' that applies to a single parameter or that has the general meaning of ``transformation with time''.

The groups define separate regions on the global fundamental plane, not across its thickness. In fact, each group shows its own fundamental plane, which is more loosely defined for less diversified
groups. We conclude that the global FP is not a bent surface, but made of a collection of several groups characterizing several fundamental planes with different thicknesses and orientations in the parameter space. In addition, since all groups are present in all galaxy clusters, we conclude that our classification, not the FPs, is universal within the redshift range of the sample (0.007 -- 0.053).

By design in the cladistic analysis, each group supposedly gathers objects sharing a similar history. They are also related by evolutionary relationships that represent a diversification scenario. By rooting the scheme with the group of least metallicity, we find that the two most diversified groups (C6, C7) have the thinnest FP, together with an intermediate one (C3). It probably indicates that the level of diversity is linked to the number and the nature of transformation events like collapse, accretion, interaction and merging, dry and wet, and that the fundamental plane is the result of several such transforming events.

Our groups have distinct multivariate properties that can thus be interpreted in the light of our current knowledge and understanding of galaxy evolutionary processes. Three groups (C1, C2 and C4) probably did not form by major mergers and have been strongly affected by interaction, some of the gas in C2 objects having possibly been swept out. Three other groups, C3, C6 and C7 have been formed by dissipative (wet) mergers because they follow a tight FP relation. In C3, this(ese)) merger(s) must have been quite ancient because of the relatively low metallicity of its galaxies. But contrarily to C7, both C3 and C6 have subsequently undergone dry mergers to increase their masses, more violent in C6 than in C3. Also, in the k-space, the C7 group clearly occupies the region where bulges (of lenticular or spiral galaxies) lie. It could have formed through minor mergers and accretions.

It has been recognized that the properties of the FP depends much on the sample. Our approach strongly associates the ``sample'' with the ``evolutionary group''. 
Since galaxies are evolutive objects, homology, a concept often used for FP studies, can be more rigorously defined
by 'similarity due to same class of progenitor'. Since galaxy histories are so complex, only multivariate studies can objectively construct homologous groups. Cladistics is indeed designed to build homologous groups and provides an evolutionary scenario that relates them. It must be understood that the assembly and interaction histories of nearby galaxies are necessarily complex and comprises often several significant transforming events \citep[e.g.]{jc2}. The interpretation, based on specific assembly histories of galaxies,  of our seven groups shows that they are
truly homologous. Our work also shows that multivariate cluster analysis is able to find homologous groups even though it cannot predict the evolutionary relationships.

The properties of the 7 groups clearly reveal that they differ in assembly histories. Since they have been obtained directly from several observables, the interpretation of the result does not depend on any a priori classification. In particular, it partly matches refined morphological classification because we used 3 parameters (among 4) that are linked to structural properties of galaxies. However our classification is more easily compared to models and numerical simulations. Our work can be readily repeated with additional observables.

In addition, the diversification scenario relating these groups does not depend on models or numerical simulations. The astrocladistic analysis was based on the assumption that the four parameters $\log{R_e}$, $<\mu_e>$, $\log\sigma$ and
$Mg_2$ are evolutive characters, that evolve and can characterize states of evolution of galaxies. The astrophysical interpretation of the diversification scenario and of the evolutionary groups demonstrates a posteriori that this assumption is correct.
This is another proof that cladistics can be applied in astrophysics. Since both cluster and cladistic analyses are only
valid for the sample in study and the variables used, the present study will be extended to other galaxy samples with more parameters.

\section*{Acknowledgments}

We thank the anonymous referee for useful comments. This research has made use of the NASA/IPAC Extragalactic Database (NED) which is operated by the Jet Propulsion Laboratory, California Institute of Technology, under contract with the National Aeronautics and Space Administration. 


\appendix

\section[]{Cluster analysis}
\label{appendClus}

Cluster analysis is the art of finding groups in data. Over the last forty years
different algorithms and computer programs have been developed for
cluster analysis. The choice of a clustering algorithm depends both on the type
of data available and on the particular purpose. Generally
clustering algorithms can be divided into two principal types viz.
partitioning (K means) and hierarchical methods.

 A partitioning method constructs K clusters i.e. it classifies
 the data into K groups which together satisfy the requirement of
 a partition such that each group must contain at least one object and
 each object must belong to exactly one group. So there are at
 most as many groups as there are objects ($K <=n$). Two different
 clusters cannot have any object in common and the K groups
 together add up to the full data set. Partitioning methods are
 applied if one wants to classify the objects into K clusters
 where K is fixed ( which should be selected optimally). The aim
 is usually to uncover a structure that is already present in the
 data. The K- means method of  \citet{MacQueen1967} is probably the
 most widely applied partitioning clustering technique.

  Hierarchical algorithms do not construct a single partition with K
  clusters but they deal with all values of K in the same run. The
  partition with $K=1$ is a part of the output ( all objects are
  together in the same cluster) and also the situation with $K=n$
  ( each object forms a separate cluster). In between all values of
  $K=1,2,3,...n-1$ are covered in a kind of gradual transition. The
  only difference between $K=r$ and $K=r+1$ is that one of the r
  clusters splits in order to obtain $r+1$ clusters or two of the
  $(r+1)$ clusters combined to yield r clusters. Under this method
  either we start with $K=n$ and move hierarchically step by step
  where at each step two clusters are merged depending on
  similarity until only one is left i.e. $K=1$ (agglomerative) or
  the reverse way i.e. start with $K=1$ and move step by step
  where at each step one cluster is divided into two (depending on
  dissimilarity) until $K=n$ (divisive). We feel that for the problem under
  consideration the partitioning method is more applicable because
  \begin{itemize}
\item[(a)] A partitioning method tries to select best clustering with K
  groups which is not the goal of hierarchical method.
\item[(b)] A hierarchical method can never repair what was done in
  previous steps.
\item[(c)]  Partitioning methods are designed to group items rather than
  variables into a collection of K clusters.
\item [(d)] Since a matrix of distances (similarities) does not have to
  be determined and the basic data do not have to be stored during
  the computer run partitioning methods can be applied to much
  larger data sets.
   \end{itemize}

For K- means algorithms \citep{Hartigan1975} the optimum value of K
  can be obtained in different ways.
In the present work the method developed by Sugar and James (2003)
has been used. This procedure is based on rate of distortion
theory. By 'distortion' we mean a measure of within cluster
variation. Let X be a p-dimensional random variable (where p
components are the parameters under consideration used for
clustering) and $C_1, C_2, ....C_K$ be a set of K cluster centers.
For simplicity, in the present study we have considered the mean
squared error as the measure of within cluster variation which is
given by $d_k $=$\frac{1}{p}min_{C_1, C_2,.....,C_K}E[(X-C_X)^{'}(X-C_X)]$
where $C_X$ is the center closest to X. Using the K means
algorithm (Macqueen 1967) we have first determined the structures
of the sub populations taking K = 1, 2, 3,..... etc. For each K we
estimated the value of distance measure
$d_{k}^{'} = \Sigma^{K}_{i=1}M_i$ where $M_i =
\frac{1}{pn}\Sigma_{j=1}^{p}\Sigma_{l=1}^{n_i}(x_{jl}^{(i)} -
\bar{x}_j^{(i)})^2$,
 $x_{jl}^{(i)} = l^{th}$ galaxy observation for
the $j_{th}$ parameter in the$ i^{th}$ cluster 
($l= 1,2,...,n_i$, j=1,2,....,p) ,
 $\bar{x}_j^{(i)}$ = average
value for the $j^{th}$ parameter in the$ i^{th}$ cluster, p is the
number of parameters, n is the number of galaxies and $n_i$ is the
number of galaxies in the $i^{th}$ cluster (i=1,2,....,K).
 Here
$\bar{x}_{j}^{(i)}$ is the estimate of the cluster center closest
to the observations in that cluster. Here $d_{K}^{'}$ can be
considered as the estimated minimum achievable distortion
associated with fitting K centers to the data. A natural way of
choosing the number of clusters is to plot $d_{K}^{'}$ versus K
and and look for the resulting curve (known as the distortion
curve). This curve is always monotonically decreasing. (Initially
we would expect much smaller drops for K greater than the true
number of clusters because past this point          ).
According to Sugar and James (2003) , the distortion curve when
transformed to an appropriate negative power, will exhibit a sharp
jump at the true number of clusters. In fact they have proved
\citep[equation (7) of ][]{sugar2003} that
   $d_K^{-p/2} \sim aK/G$ for  $K\ge G $ and 
                    $\sim 0$ for $K<G$,
where G is the actual number of clusters and $0<a<1$.\\
The above relation suggests several possibilities for determining
the actual number G. In particular one can use the jump method which
estimates G using 
$arg max_{K}[{d_{K}^{'}}^{-y} - {d_{K-1}^{'}}^{-y}]$,
the value of K associated with the largest jump in the transformed
distribution. It also suggests that an appropriate value of y
would be p/2. In our case as the number of parameters is 3 and 4,
one can take y as 1.5 and 2 respectively.

\begin{figure}
   \centering
 \includegraphics[width=\columnwidth]{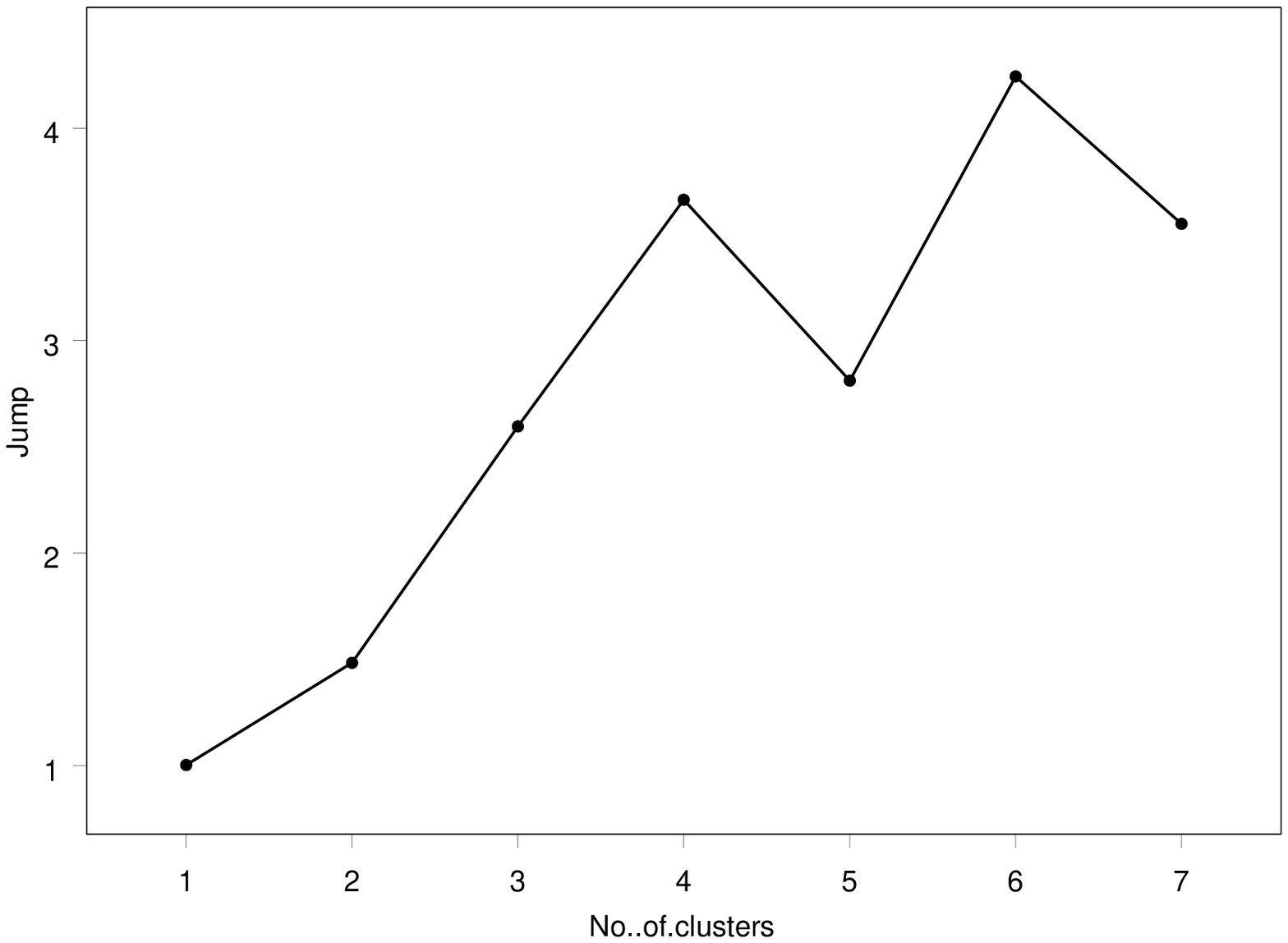}
 \includegraphics[width=\columnwidth]{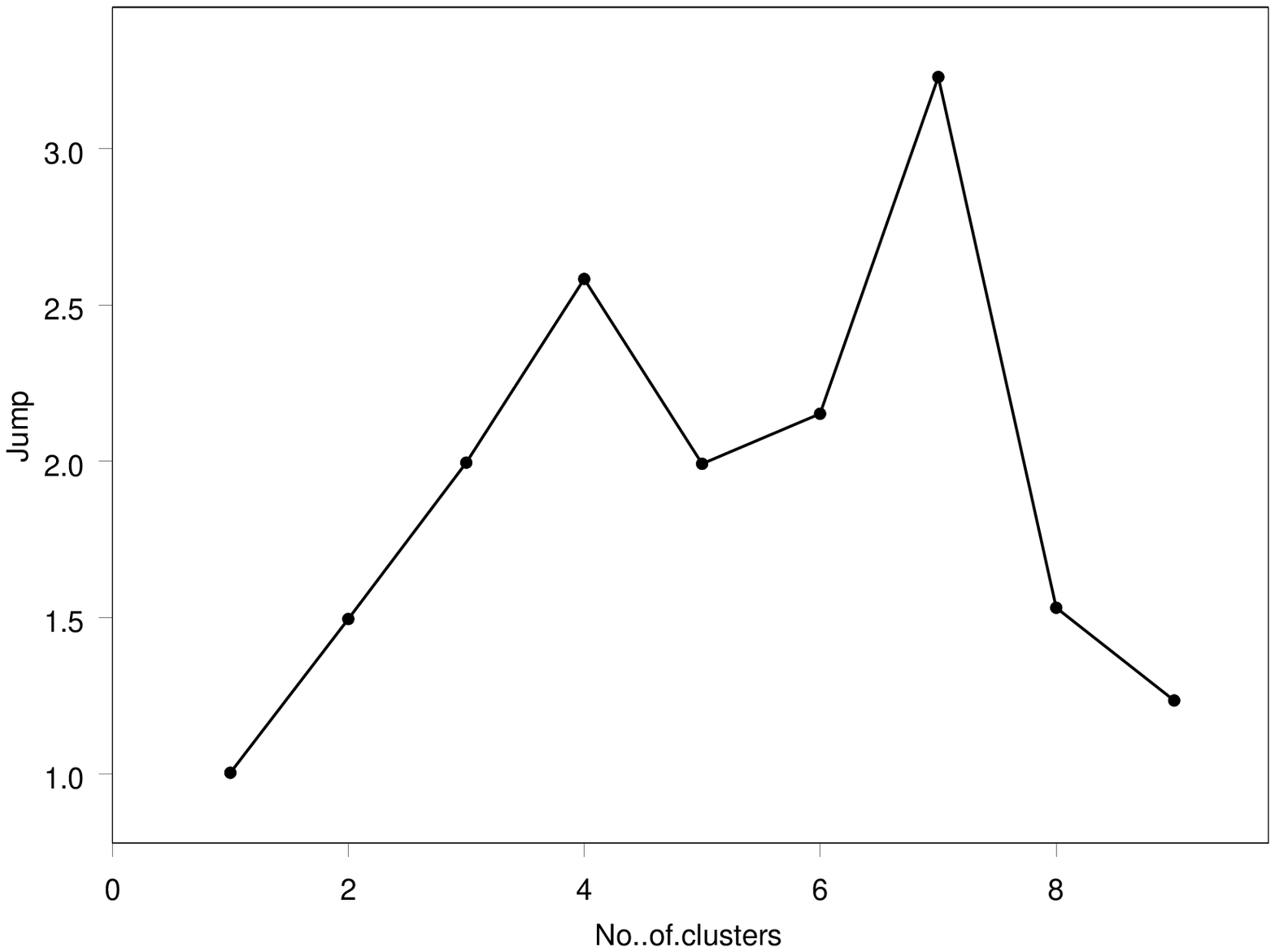}
   \caption{Jump $J_{K}$ as a function of $K$ for the 3- (top) and 4- (bottom) parameter cases.} 
    \label{figjump}%
    \end{figure}

We have calculated
the jumps in the transformed distortion as
 $J_{K} = (d_{K}^{\bf'-2}  - d_{K-1}^{\bf'-2}$).
The optimum number of clusters is the value of  K associated with
the largest jump. The largest jump can be determined by plotting
$J_{K}$ against K and the highest peak will correspond to the
largest jump. In the present situation there are jumps at K=4 and
K=6 for 3 parameter case and K=4 and K=7 for 4 parameter case  (Fig.~\ref{figjump}). The
largest jumps are at K=6 and K=7 in the former and latter cases
respectively.

\section[]{Cladistic analysis}
\label{appendClad}

\begin{figure}
   \centering
 \includegraphics[width=\columnwidth]{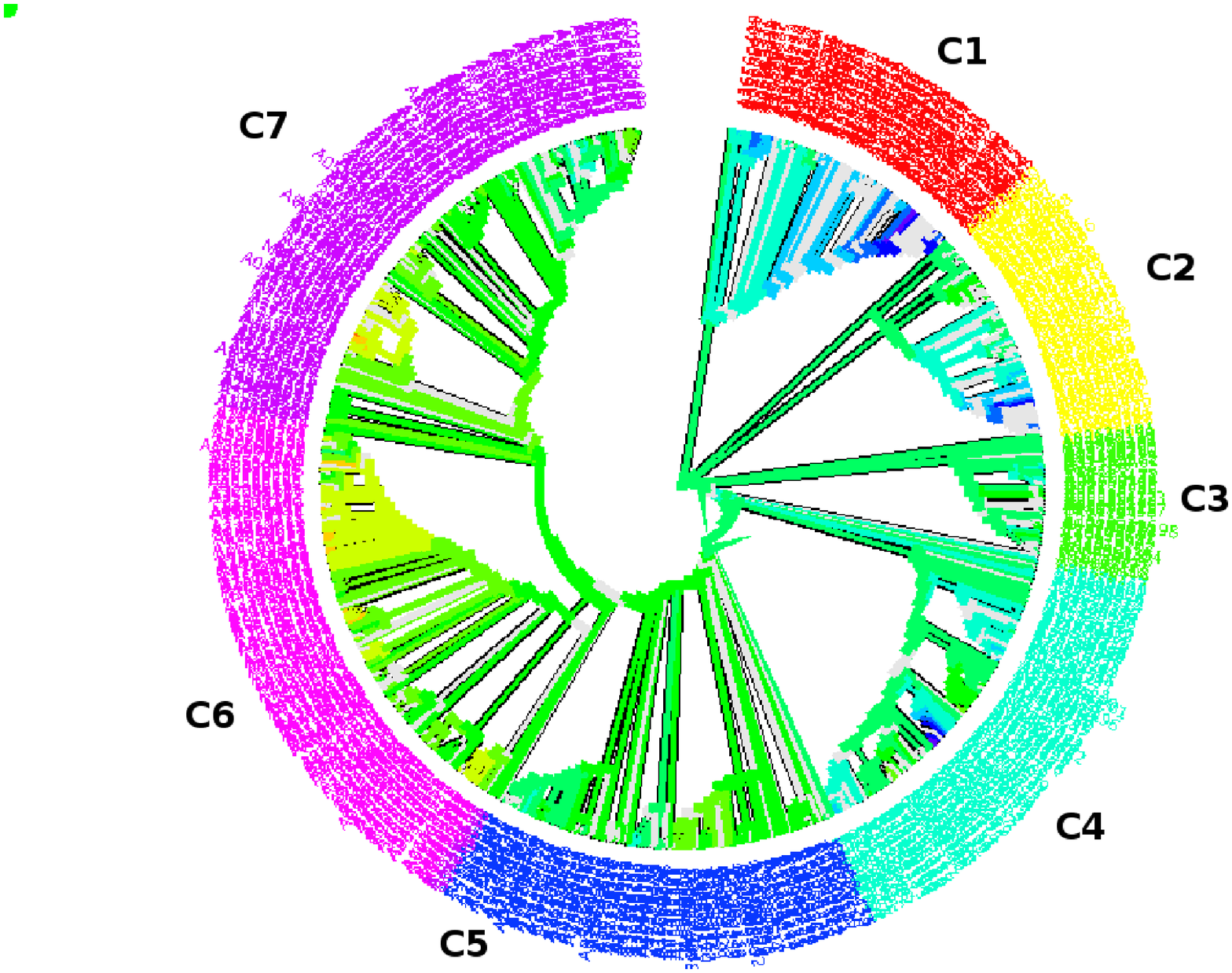}
   \caption{Cladogram obtained with  4 parameters. It is a majority rule consensus of 986 equally parsimonious trees.
The branches are coloured according to the value of Mg2 (from dark blue to yellow through green as Mg2 increases). The outer annulus is formed by the names of the galaxies, with colours
according to the group (numbered from C1 to C7) as detailed in Table~\ref{tabgalaxiesgroups}.
} 
    \label{cladograms}%
    \end{figure}

Cladistics is the subject of numerous books \citep[e.g.][]{compleat1991, semple2003} and the astrocladistic methodology has been presented in detail in several papers \citep{FCD06,jc1,jc2,DFB09,FDC09}. We refer the reader to these references for a complete description of the method used here.
For the present analysis, we have built a matrix with values of the four parameters for all galaxies. The values for each parameter were discretized into 30 equal-width bins representing
supposedly evolutionary states \citep[see e.g.][]{Goloboff2006,TF09}. We adopted the popular parsimony criterion, which selects the most parsimonious tree among all possible arrangements because it represents the simplest evolutive scenario for the sample. (Maximum parsimony corresponds to the minimum number of changes of states for all parameters that occur along the paths between all objects. This number is unique to each tree.) Fig~\ref{cladograms} is a majority-rule consensus tree reflecting the most common features in all equally most parsimonious trees. For this paper, the maximum parsimony searches were performed using the heuristic algorithm implemented in the
PAUP*4.0b10 \citep{paup} package, with the Multi-Batch Paup Ratchet method (http://mathbio.sas.upenn.edu/mbpr). Heuristic methods do not explore the parameter space of all possible tree arrangements, which would take a prohibitive computer time for hundreds of objects, but try to find the minima. They cannot guarantee finding the absolute most parsimonious trees
but generally require far less computer time while being quite effective. The Ratchet approach still improves this efficiency.
The results were interpreted with the help of the Mesquite software
\citep{mesquite} and the R-package (used for graphics and statistical analyses).
In contrast to multivariate distance methods, undocumented values are not a problem in cladistic
analyses. This is why the galaxies missing $Mg_2$ determination have not been excluded.

In the two cases of the present study, with and without $Mg_2$, we find rather robust trees. To assess this robustness, we
proceeded in the following way. We first selected 3 clusters, A1656 (Coma cluster, 56 galaxies, the largest cluster in this sample), A2199 (38 galaxies) and A3526 (37 galaxies). Cladistics analyses
were run on these 3 subsamples separately, and then with the 3 together. The four tree structures were compatible, indicating that an underlying arrangement did exist. This was confirmed by the
analysis done on the whole sample that converged quite easily despite the low number of characters as compared to the number of objects. It must be noted that the arrangement of the objects on a
tree is constrained by the level of information provided by the parameters. When their number is relatively low, the constraints are weaker and the number of possible most parsimonious trees is
larger, making the convergence toward compatible trees less probable. In the present case, we find that the 3-parameter result is slightly less robust. Nevertheless, all trees with and without $Mg_2$ are compatible.

The choice of the root of the tree orientates the evolutionary processes and the ``rank'' of diversification of the groups as seen on Fig~\ref{cladograms}. It indicates the ancestral states of the characters. The trees of the present study are rooted with objects or group
of objects having low $Mg_2$ because we find it is the only objective criterion of ancestrality. However, this choice must imply a consistent evolutionary scenario for all the other parameters, which is the case here.

 \section[]{Complementary information}
\label{otherfigures}

\begin{figure}
   \centering
 \includegraphics[width=\columnwidth]{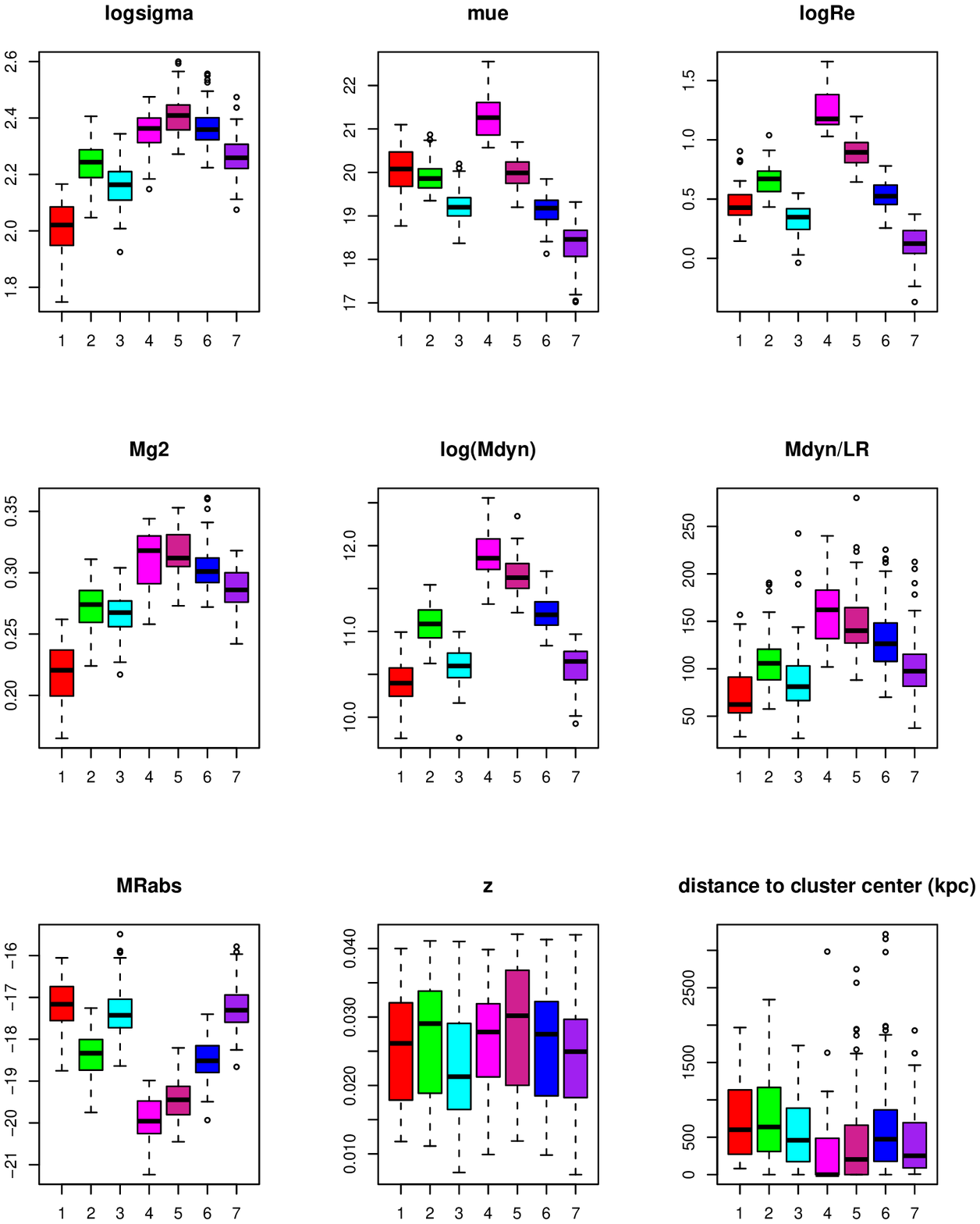}
   \caption{Same as Fig.~\ref{boxplot} but for groups from cluster analysis with 4 parameters and K=7. The colours are the same as in Fig.~\ref{gpclust} (middle row, right column).} 
    \label{C4K7boxplot}%
    \end{figure}

We present here three figures for the other groupings found by the other analyses. Fig.~\ref{C4K7boxplot} is the equivalent of Fig.~\ref{boxplot} but for the cluster analysis with 4 parameters and
K=7 with colours as in Fig.~\ref{gpclust} (middle row, right column). If ones recall that C6 and C7 are split into 4 groups (4 to 7 on the figure), and that C2 and C4 are mixed together in this
cluster analysis, the correspondence between the two boxplot figures is striking.

\begin{figure*}
    \centering
 \includegraphics[width=5 cm]{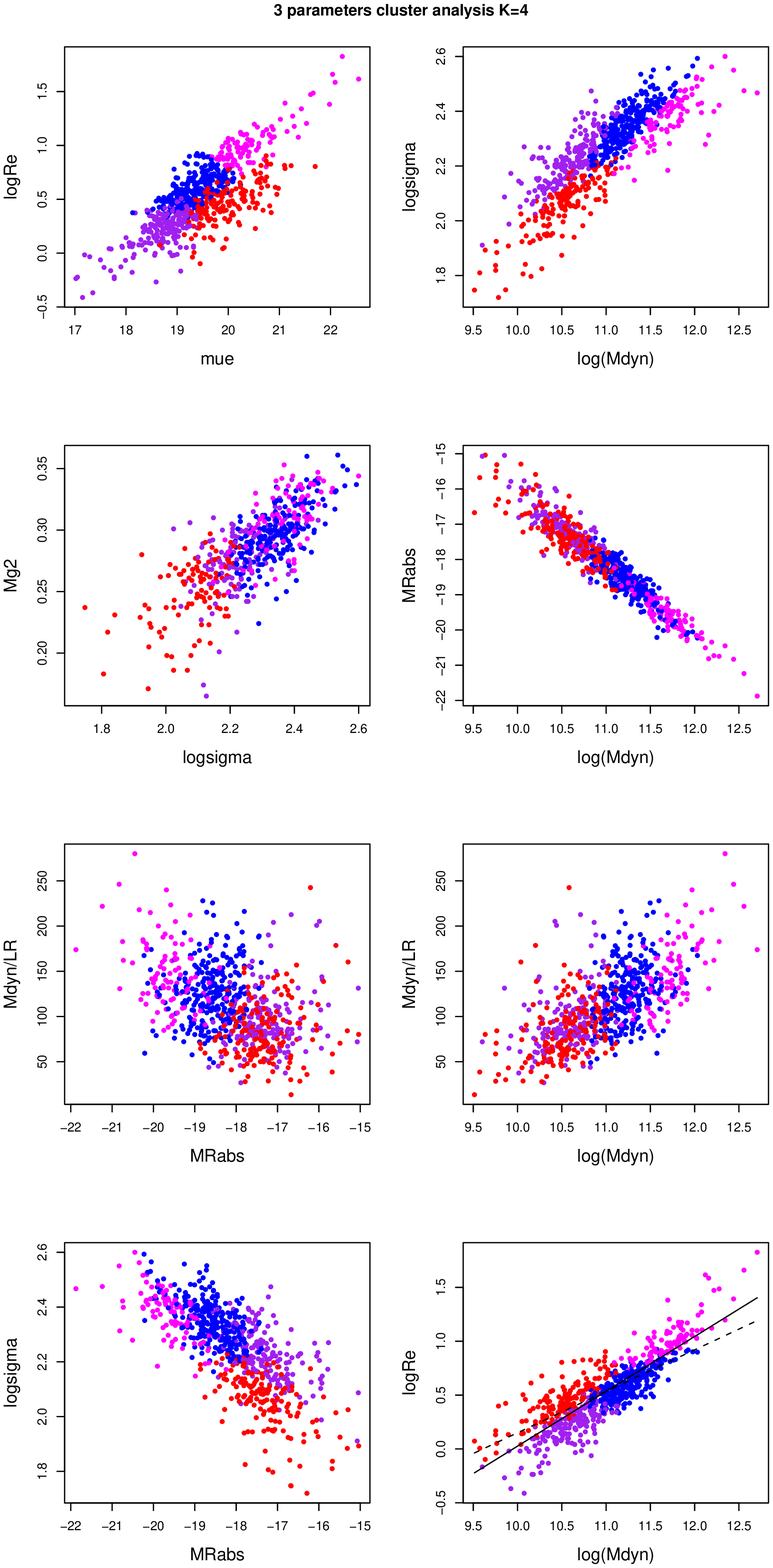} \hskip 1 true cm
 \includegraphics[width=5 cm]{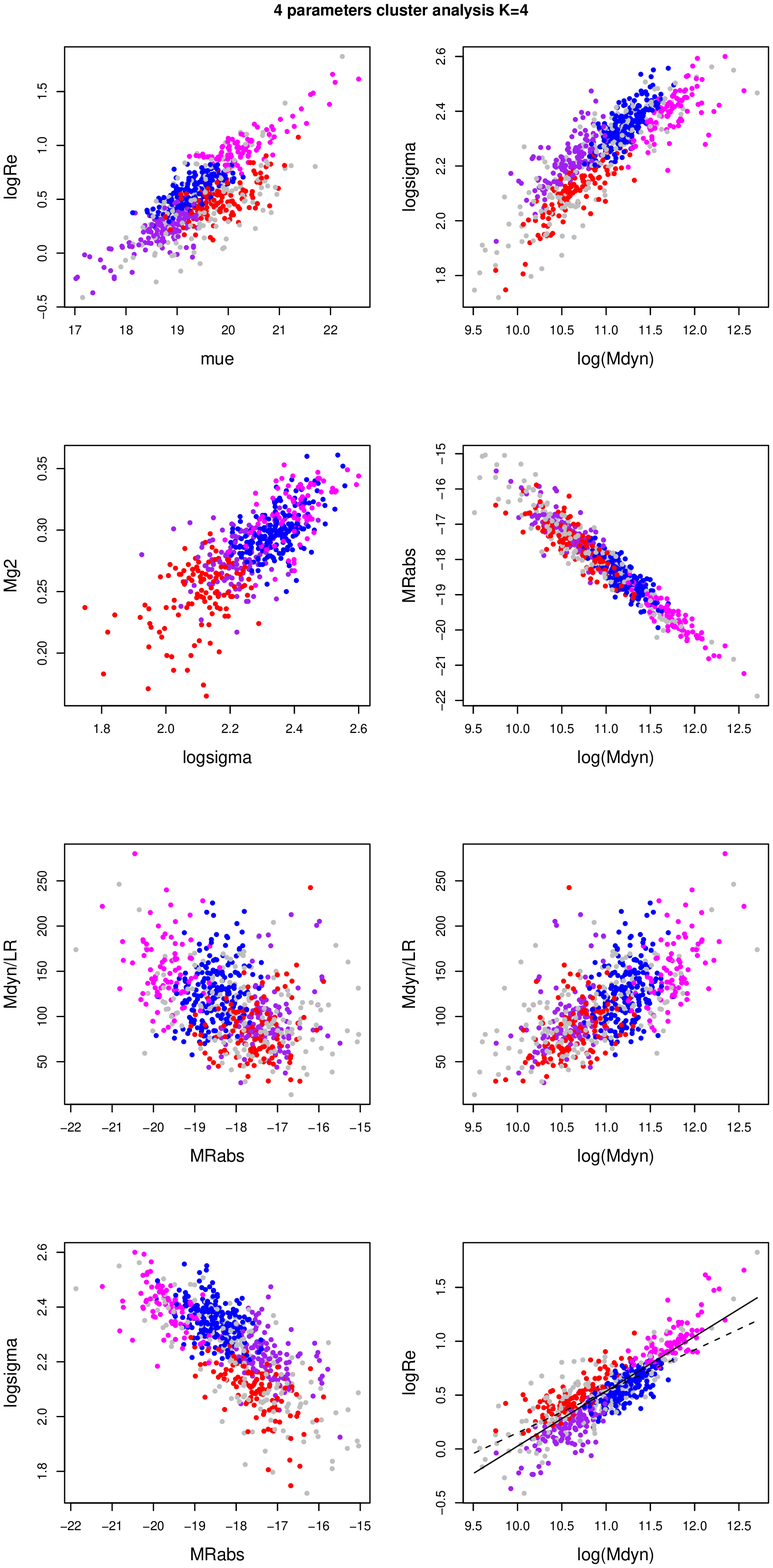}
   \caption{Diagrams as in Fig.~\ref{figcorr1}, but for the cluster analyses with K=4.The colours are the same as in Fig.~\ref{gpclust} (top row).} 
     \label{figcorralltop}%
     \end{figure*}

\begin{figure*}
    \centering
  \includegraphics[width=5 cm]{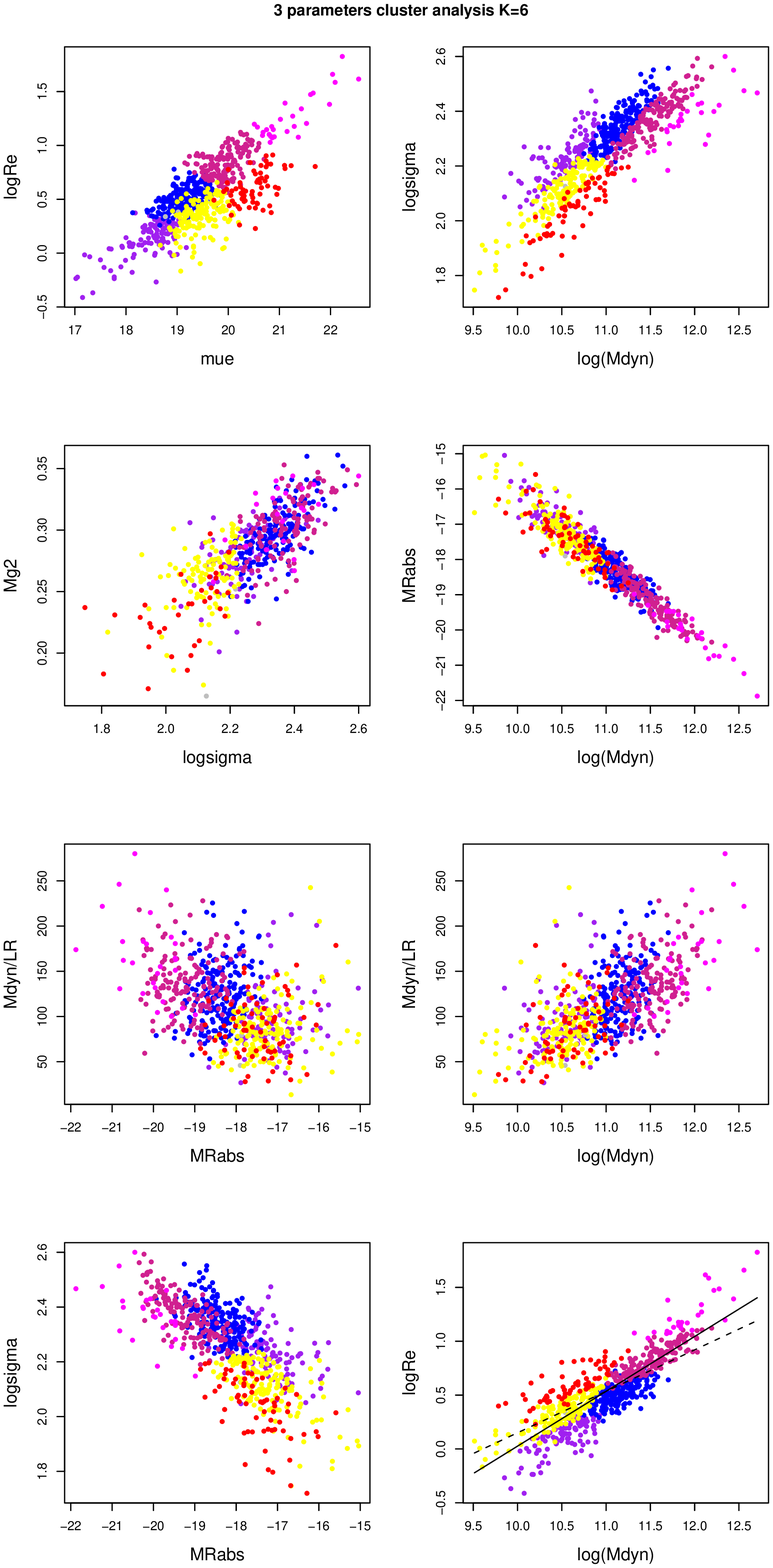} \hskip 1 true cm
 \includegraphics[width=5 cm]{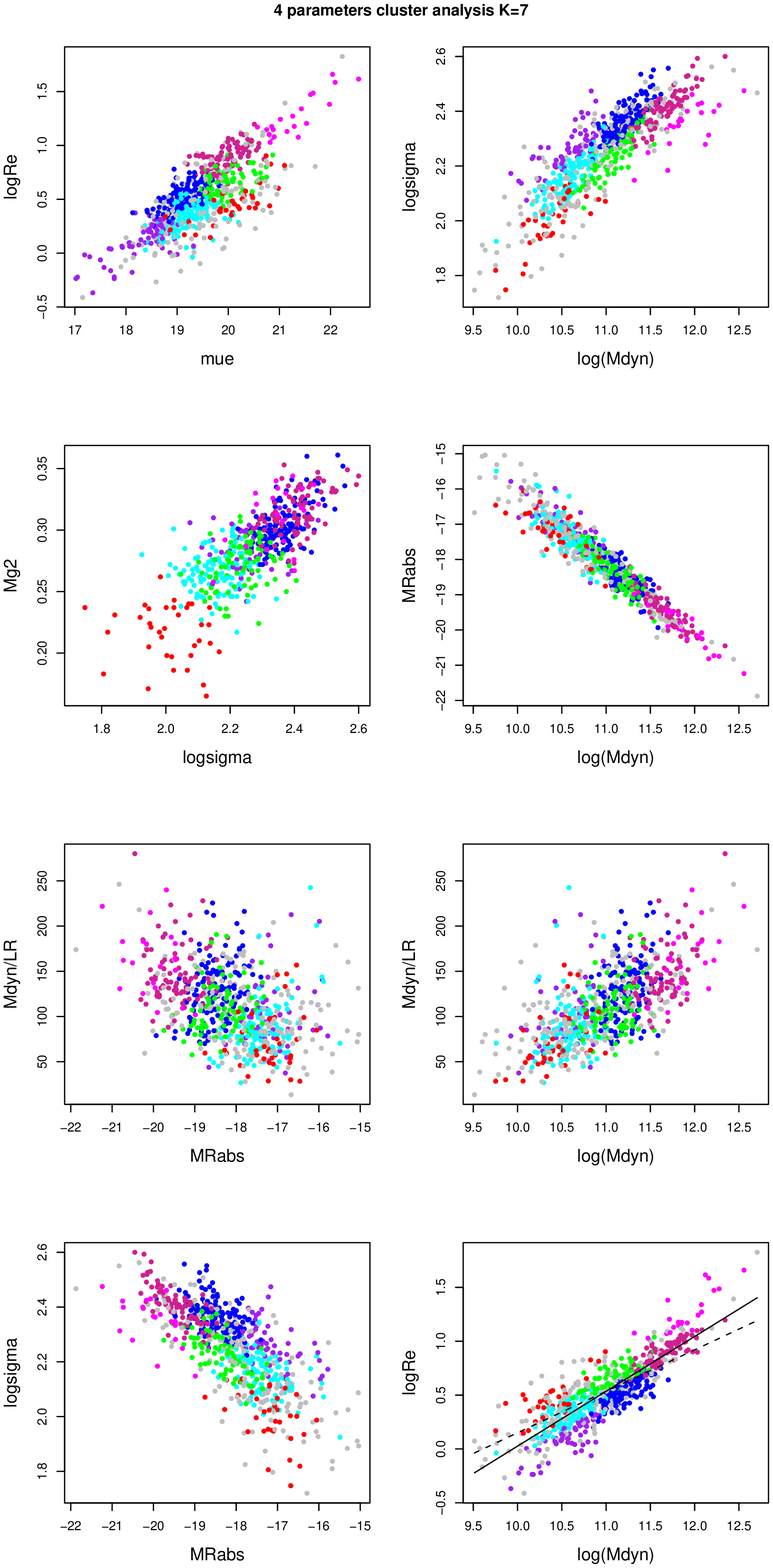}
    \caption{Diagrams as in Fig.~\ref{figcorr1}, but for the cluster analyses with K=6 and K=7.The colours are the same as in Fig.~\ref{gpclust} (middle row).} 
     \label{figcorrallmiddle}%
     \end{figure*}

\begin{figure*}
    \centering
  \includegraphics[width=5 cm]{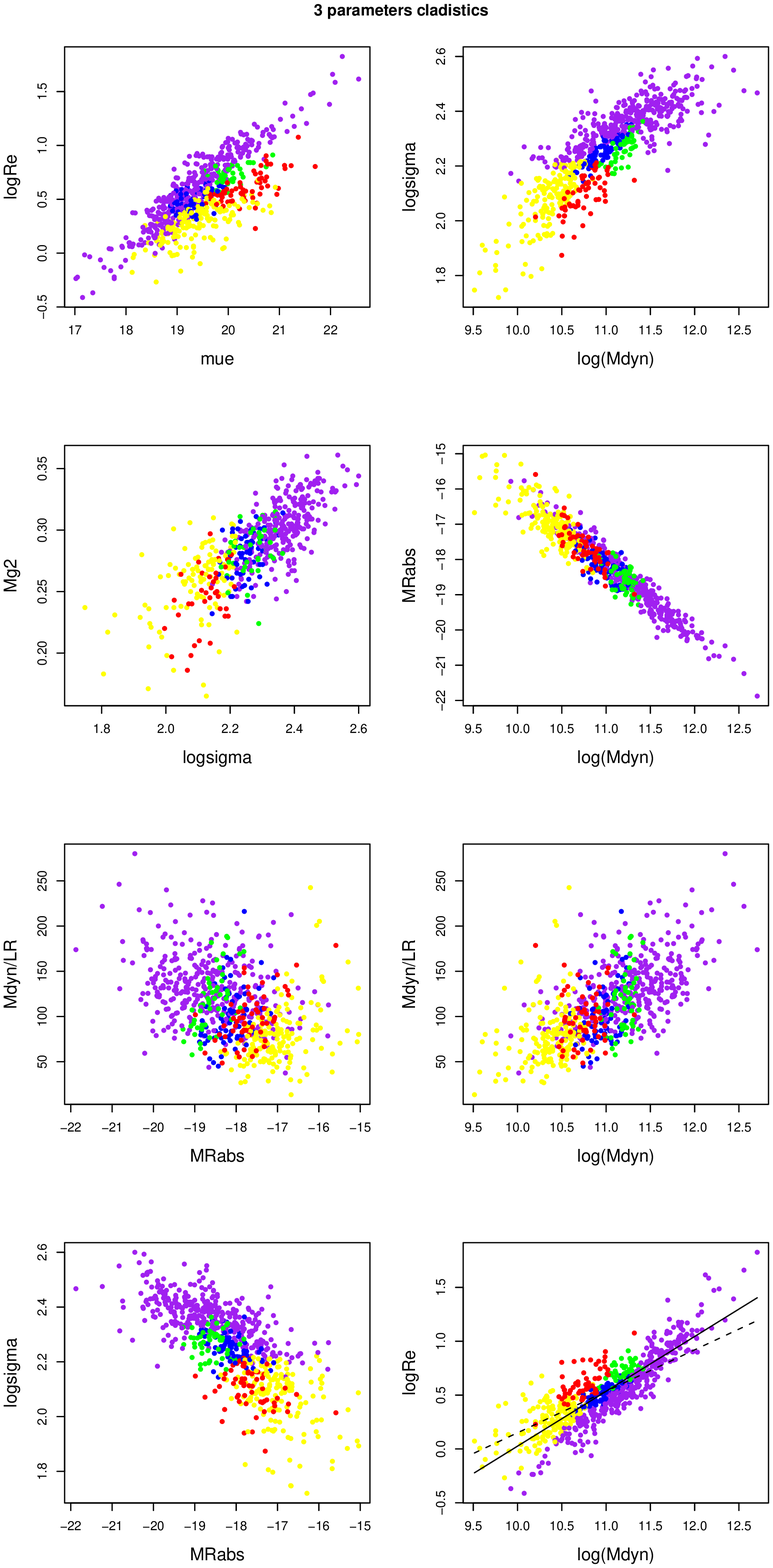} \hskip 1 true cm
   \includegraphics[width=5 cm]{fig7_correl.eps}
    \caption{Diagrams as in Fig.~\ref{figcorr1}, for the two cladistic analyses with 3 and 4 parameters. The colours are the same as in Fig.~\ref{gpclust} (bottom row).} 
     \label{figcorrallbottom}%
     \end{figure*}

\begin{figure*}
   \centering
   \includegraphics[width=6 cm]{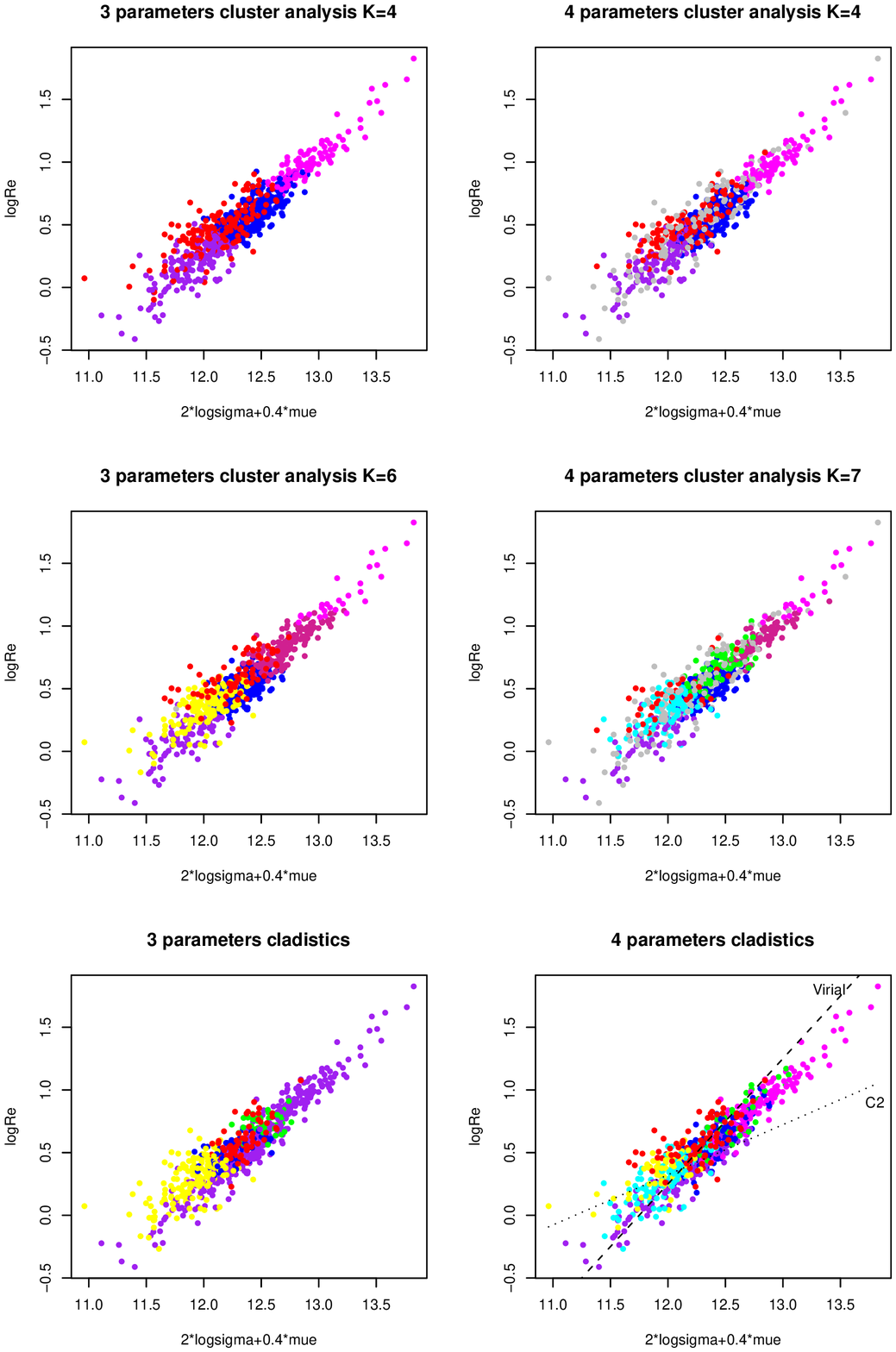}
 \caption{Tilt with respect to the virial plane like in Fig.~\ref{figFPtilt} for all the groupings.}
    \label{figFPtiltall}%
    \end{figure*}

Fig.~\ref{figcorralltop}, Fig.~\ref{figcorrallmiddle} and Fig.~\ref{figcorrallbottom} are the equivalent of Fig.~\ref{figcorr1} for all the analyses.
Fig.~\ref{figFPtiltall} shows the tilt with respect to the virial plane like in Fig.~\ref{figFPtilt} for all the groupings.

Finally, we give in Table~\ref{tabgalaxiesgroups} the list of all the galaxies used in this study distributed in the groups C1 to C7.

\newpage

\begin{table*}
 \begin{minipage}{18cm}
  \caption{List of galaxies for each group.}
     \label{tabgalaxiesgroups}
\begin{tabular}{lllllll}
\hline
 \multispan{7} C1              \\
\hline
A2199:B-008     &  A1060:ST-026    & E323-008        &  A3558:FCP-35    &  I1618           & A2052:EFR-B     &  U01269           \\
A0576:SMC-A     &  A1177:SMC-B     & A3733:SMC-K     &  A3574:W-050     &  PISC:PP-Z10020  & A2063:D-033     &  S0301:D-022       \\
N4661           &  A1656:D-107     & N0085A          &  A2052:MKV-13    &  A0194:D-071     & I4767           &  A0539:D-064       \\
A1656:D-204     &  A3558:FCP-31    & E243-052        &  A2063:D-034     &  S0301:D-034     & A4038:D-067     &  A0957:D-035        \\
7S21:PP-S06     &  S0761:FCP-14    & N0548           &  A2199:B-034     &  A0539:D-054     & A4038:D-037     &  A1060:JFK-R225      \\
A2877:D-040     &  A2063:D-035     & S0301:D-024     &  A4038:D-083     &  A0957:D-049     & A0576:SMC-B     &  A1139:D-016       \\
A0189:SMC-I     &  A2199:B-087     & A0426:PP-P07    &  A4038:D-039     &  A1016:SMC-G     & A1016:SMC-A     &  A3526:D-047       \\
A0262:PP-A05096 &  A2634:D-087     & A3381:D-037     &  U02717          &  A1060:JFK-RMH50 & A1656:D-081     &  A3558:FCP-39       \\
I0293           &  A4038:D-052     & A1016:SMC-F     &  A1016:SMC-E     &  A3526:D-026     & E445-040        &  S0761:FCP-24       \\
A0539:D-043     &  A0262:B-042     & A1060:JFK-R245  &  I4041           &  A3558:FCP-56    & A2877:D-033     &  A2063:D-065         \\
A0957:D-033     &  A0576:SMC-J     & E322-100        &  A4038:D-045     &  E445-054        & A0189:SMC-J     &  A2199:RS-028         \\
A3733:SMC-G     & A4038:D-033     &   S0761:FCP-07    &       \\                                                                                   
\hline
  \multispan{7}  C2            \\
\hline                                                                                                                                             
7S21:PP-S07     & A1656:D-087    & A0569:SMC-L     &  A2806:SMC-D   & I4011               &  A1060:JFK-R338        &  A3389:D-053    \\                                                                 
                                    
N0386           & A3558:FCP-15   & A1060:JFK-RMH35 &  N0398         & A3571:SMC-40        &  A3526:D-040           &  A0999:SMC-C     \\                                                                
                                     
A2877:FCP-24    & E384-029       & A1177:SMC-C     &  I1638         & S0753:W-095         &  A3526:D-027           &  A3526:D-059     \\                                                                
                                     
H0122:PP-H01051 & S0805:FCP-09   & E322-102        &  A0194:D-052   & A3716:D-061         &  A3526:D-015           &  A1656:D-135      \\                                                               
                                      
A0262:B-038     & A4038:D-076    & N4729           &  S0301:D-026   & A4038:D-053         &  A1656:D-156           &  A4038:D-066       \\                                                              
                                       
S0301:D-031     & A0957:D-050    & Z160-027        &  A3389:D-043   & PISC:PP-Z01034      &  A1656:D-027           &  S0761:FCP-26    \\                                                                
                                     
N2330           & A2806:SMC-F    & A1656:D-193     &  A0957:D-037   & A2877:D-011         &  S0753:W-017           &  S0301:D-027     \\                                                                
                                   
A1060:JFK-RMH26 & PISC:PP-Z01032 & A3570:SMC-64    &  E501-049      & I1648               &  A2199:B-044           &  A0400:D-057      \\                                                               
                                    
E437-045        & A2877:D-025    & S0753:W-051     &  A1257:SMC-GC  & U01040              &  A4038:D-060           &  A3526:D-035      \\                                                               
                                       
E322-099        & A0194:D-045    & E104-002        &  A3526:D-050   & S0301:D-020         &  A4038:D-049           &  N4743             \\                                                              
                                        
\hline
  \multispan{7}  C3          \\
\hline                                                                                                                                             
  N0420          &    A2199:B-015    & A0569:SMC-R    & N2235          &  N5304          &  A0548:D-007    &  I3955          \\           
  A0426:PP-P08   &    A0347:PP-B16   & N4616          & J8:PP-J03049   &  A4038:D-043    &  A0957:D-043    &  A0539:D-031     \\          
  A3381:D-112    &    A3733:SMC-I    & A3571:SMC-10   & A0539:D-059    &  A0539:D-042    &  A3558:FCP-57   &  I0310            \\         
  A1257:SMC-G    &    I0171          & A2634:D-031    & N2329          &  A3716:D-098    &  MKW12:FCP-09   &  A3381:D-025    \\           
  A3570:SMC-50   &    A0539:D-039    & A0426:PP-P11   & N4929          &  I1860          &  A4038:D-065    &  A0957:D-044    \\           
  A3558:FCP-17   &    I1116          & A0262:B-018    & A3571:SMC-44   &                 &                 &                  \\   
\hline
  \multispan{7}  C4        \\
\hline                                                                                                                                             
 I1548          &  J8:EFR-H         & S0753:W-010     & A1736:D-144     & A3571:SMC-29  &   A3381:D-033  &   A1257:SMC-C   \\
 A2877:D-035    &  A0426:PP-P33     & A2199:B-054     & S0761:FCP-05    & E286-029      &   A1314:SMC-E  &   E322-089       \\
 A0539:D-041    &  A1060:JFK-RMH28  & A3656:SMC-X     & A2199:B-047     & A2806:SMC-E   &   A1656:D-210  &   N4824          \\
 A1016:SMC-C    &  A1314:SMC-D      & N0909           & S0805:D-021     & A2877:D-021   &   I3957        &   N4906           \\
 A3526:D-049    &  A3526:D-036      & E488-009        & A2634:L-BO3C    & I1696         &   A3581:SMC-76 &   A3558:FCP-29     \\
 A1656:D-238    &  A1656:D-153      & A1257:SMC-E     & N0388           & A0400:D-017   &   A2199:B-095  &   A2199:B-028    \\
 N4876          &  I4133            & A1656:D-096     & U01003          & A3389:D-048   &   A4038:D-068  &   I4748          \\
 A2199:B-048    &  S0753:W-047      & A3716:D-081     & A3381:D-067     & A1177:SMC-H   &   A0539:D-051  &   A2634:D-079     \\
 A2199:B-073    &  A2199:B-033      & A4038:D-059     & A1060:ST-034    & E322-075      &   A0569:SMC-B  &   N0501           \\
 U01030         &  S0805:D-029      & PISC:PP-Z01073  & A3526:D-033     & E323-009      &   I2955        &   N4850            \\
 A0548:D-020    &  A2634:D-104      & A0194:D-012     & I3947           & A1656:D-207   &   E384-036     &   N4882             \\
 E437-021       &  PISC:PP-Z01047   & J8:PP-J01080    & S0753:W-012     & A1736:D-039   &   A2634:B-013  &   A1314:SMC-B    \\
 A3537:SMC-156  &  U00996           & A0539:D-052     & A2199:B-084     & A2063:D-046   &   A2877:D-037  &   A0426:PP-P21    \\
 E510-054       &  A0539:D-075      & A1139:D-041     & A3656:SMC-S     & A2199:B-021   &   I1680        &   A0957:D-046     \\
 A2634:D-093    &  E436-044         & A1367:B-020     & A0539:D-063     & E235-039      &   A0262:B-019  &   A0426:PP-P20     \\
 N0397          &  A3526:D-009      & E322-101        & I0464           & A2877:D-048   &                &                    \\
 \hline
\end{tabular} 
\end{minipage}
\end{table*}

\begin{table*}
 \begin{minipage}{18cm}
\contcaption{}
 \begin{tabular}{lllllll}
\hline
   \multispan{7} C5            \\
\hline                                                                                                                                             
 N0394          & U02673         & A3581:SMC-75   & A0539:D-044    & A2063:D-050    & I5354          & A3558:FCP-33     \\        
 N0385          & I2744          & A2063:D-089    & A3389:D-060    & A3716:D-117    & N0759          & E445-059         \\       
 N0712          & N4946          & A3656:SMC-I    & A1016:SMC-B    & N0912          & I1858          & A2063:D-073       \\       
 A0426:PP-P15   & N5424          & A3744:SMC-I    & N4767          & A2063:D-071    & A0576:SMC-I    & A2199:B-026        \\       
 A3381:D-075    & A2199:B-005    & A2634:D-068    & N4908          & A2877:D-045    & N4881          & A3733:SMC-H         \\       
 A0957:D-054    & A2634:B-021    & I1569          & E325-013       & A0400:D-070    & A3558:FCP-21   & A2634:D-038       \\       
 N3841          & A0076:D-016    & J8:EFR-I       & E511-021       & A0539:D-062    & A2199:EFR-O    & N0079           \\       
 A1656:D-161    & A3526:D-041    & A0539:D-057    & N6146          & A0569:SMC-N    & A3733:SMC-B    & U01837             \\       
 A3558:FCP-06   & A2877:D-028    & N4816          & A3656:SMC-P    & E501-003       & A0999:SMC-G    & N1224              \\       
 A3581:SMC-78   & J8:PP-J07038   & A1736:D-137    & A2634:D-043    & A1656:D-140    & N0382          & N3308               \\       
 A2063:D-059    & N1273          & N5438          & A2657:D-070    & A1656:D-230    & N0560          & A1656:D-206          \\       
 A2199:EFR-H    & A3381:D-100    & A3716:D-090    & N0541          & A3571:SMC-164  & A0400:D-089    & A3574:W-024       \\       
 A3744:SMC-T    & A0999:SMC-E    & A0189:SMC-C    & I1806          & A2052:MKV-60   & A0539:D-016    & N6158              \\       
 A2634:D-107    & E322-081       & A2063:D-074    & A3381:D-034    & A2199:B-019    & I0458          & A3744:SMC-Q        \\       
 A0076:D-018    & N4864          & A2877:D-042    & N4854          & A3716:D-099    & A1139:D-030    & N4683              \\       
 J8:EFR-K       & A3558:FCP-04   & I1807          & A3558:FCP-07   & A2634:D-130    & A1656:D-239    &                       \\
 \hline           
 \multispan{7}   C6        \\
\hline                                                                                                                                             
A2806:SMC-C   & N6998         & N3309         & A2063:D-072   & N2340         & E325-004      & N0410            \\        
A0400:D-058   & I5341         & A1314:SMC-A   & A2634:D-074   & N4696         & E510-066      & N0564             \\       
A0539:D-045   & A2657:D-031   & A1656:D-240   & A4049:D-055   & E444-046      & N6173         & N1278             \\       
A0957:D-059   & N0212         & E509-008      & I1566         & N6999         & A3716:D-080   & A0999:SMC-D        \\      
S0753:W-073   & A0539:D-047   & A3558:FCP-13  & E243-045      & N0703         & E464-018      & I2738               \\     
A3716:D-067   & I0661         & I4329         & N0508         & I0313         & N7728         & N3862             \\       
A2634:B-016   & N4889         & E511-026      & N1272         & A0569:SMC-Q   & I5353         & N4926             \\       
N0083         & N6160         & E187-020      & N2230         & N4869         & J8:EFR-D      & A3558:FCP-03       \\      
N0379         & E243-049      & A3733:SMC-C   & A1139:D-039   & A2199:B-020   & U05515        & E325-016           \\      
A0189:SMC-A   & I0312         & A2634:D-057   & A1314:SMC-G   & I5342         & N4839         & I4374               \\     
U01308        & A0539:D-050   & A2657:D-071   & I4051         & A4049:SMC-E   & U09799        & I4765                \\    
A0539:D-068   & I2597         & N0708         & A3558:FCP-02  & I1568         & I5358         & A3716:D-078       \\       
I0613         & A2063:D-090   & A3381:D-055   & A3558:FCP-16  & I1633         & A0347:PP-B07  & N7016              \\      
I0708         & E286-049      & N3551         & N5419         & N0545         & N1293         & N7735              \\      
N4709         & A2657:D-043   & A3558:FCP-05  & E511-032      & I1907         & A0576:SMC-D   & I5362               \\     
N4927         & I1565         & N6166         & A3716:D-065   & A0576:SMC-C   & N4923         & J8:EFR-C             \\   
A3558:FCP-18  & N0383         & N0499         & A3733:SMC-A   & I0660         & E104-007      & N3311              \\  
A3571:SMC-171 & N0507         & N1283         & A2634:D-077   & N3842         & A2634:D-036   & N4874              \\  
E511-023      & I1803         & A3381:D-021   & A2657:D-064   & E443-024      & N0080         & A2063:D-060         \\ 
I4931         & A3381:D-056   & I3959         & U01841        & A3558:FCP-08  & N0215         & S0761:FCP-04         \\
\hline
  \multispan{7}  C7           \\
\hline                                                                                                                                             
N0375             & N2332          &  A3526:D-046   &   A3558:FCP-24     &  A2634:D-056   &   I5350            &    N1270                 \\        
A0347:PP-B03C     & E437-011       &  N4875         &   A3558:FCP-14     &  A2634:D-102   &   A1139:D-029      &    U02725                 \\       
A0426:7S-PER199   & N3555          &  I4045         &   A3571:SMC-21     &  A4038:D-044   &   N0543            &    U03696                 \\       
A0539:D-061       & A1257:SMC-B    &  E384-049      &   N5423            &  N0384         &   A0400:D-041      &    N3305                   \\      
A0569:SMC-G       & N4706          &  A2199:B-035   &   A2199:B-024      &  N0528         &   A0539:D-048      &    A1177:SMC-F              \\     
A1060:JFK-RMH79   & N4919          &  A2199:B-061   &   E103-046         &  N0687         &   A3389:D-049      &    A1228:SMC-M            \\       
N4730             & A3558:FCP-25   &  A3716:D-141   &   A3716:D-084      &  A0400:D-044   &   A1060:JFK-RMH29  &    N3851                  \\       
N4872             & A3558:FCP-09   &  A2634:D-075   &   A2634:D-119      &  U02698        &   E323-005         &    N4860                   \\      
I4021             & A3571:SMC-32   &  A2634:D-069   &   A4049:SMC-D      &  A0548:D-051   &   A1656:D-136      &    A3558:FCP-34            \\      
S0753:W-049       & N5397          &  A4038:D-055   &   I1673            &  E501-013      &   I4012            &    A3558:FCP-50             \\     
A2063:D-077       & A2199:RS-008   &  N0380         &   A0400:D-052      &  A1139:D-036   &   S0753:W-037      &    A3571:SMC-38              \\    
A2199:B-074       & A2199:RS-163   &  I0103         &   A0426:7S-PER163  &  A1228:SMC-H   &   S0761:FCP-11     &    A3571:SMC-112          \\       
A2199:B-038       & A3716:D-116    &  N0679         &   A0548:D-017      &  N3837         &   A2199:L-0163     &    A2052:EFR-C             \\      
A2634:D-076       & N7014          &  J8:EFR-A      &   E436-045         &  N4840         &   A2199:B-045      &    A2199:B-030             \\      
A2634:D-071       & A4049:D-047    &  N1281         &   N4645            &  E382-002      &   A2634:D-080      &    A3716:D-051              \\     
A4038:D-070       & N3873          &  A0548:D-019   &   A1656:D-024      &  A3558:FCP-26  &   A2634:D-073      &    A3744:SMC-E               \\   
A4038:D-038       & A2877:FCP-32   &  A0999:SMC-F   &   A1656:D-125      &  E445-028      &   A2657:D-072      &    A4038:D-032             \\  
E243-041          & S0301:D-017    &  A1139:D-037   &   A3574:W-074      &  A3571:SMC-13  &   A4038:D-051      &    I0662                   \\  
N0547             & A0426:PP-P26   &  A1228:SMC-G   &   A3581:SMC-77     &  E510-063      &   N0392            &    A0539:D-049              \\ 
N0911             & A0539:D-069    &  I0709         &   A2199:B-069      &  A2199:B-050   &   N0529            &    E437-013                  \\
A0426:PP-P22      & A0957:D-047    &  E323-034      &   A2199:B-066      &  I4926         &   U01859           &    I3986                    \\
A2634:B-030       & E235-049       &                &                    &                &                    &                             \\
 \hline
\end{tabular} 
\end{minipage}
\end{table*}

\bsp

\label{lastpage}

\end{document}